\documentclass[preprint,showpacs,preprintnumbers,amssymb,nofootinbib]{revtex4}

\usepackage{graphicx}
\usepackage{amssymb}
\usepackage{epstopdf}
\usepackage{dcolumn}
\usepackage{bm}

\newcommand{\nn}{\nonumber}

\newcommand{\be}{\begin{eqnarray}}
\newcommand{\ee}{\end{eqnarray}}
\catcode`\@=11
\def\lsim{\mathrel{\mathpalette\@versim<}}
\def\gsim{\mathrel{\mathpalette\@versim>}}
\def\@versim#1#2{\vcenter{\offinterlineskip
\ialign{$\m@th#1\hfil##\hfil$\crcr#2\crcr\sim\crcr } }}
\catcode`\@=12

\def\thefootnote{\fnsymbol{footnote}}

\begin{document}

\def\thefootnote{\fnsymbol{footnote}}
\begin{flushright}
KANAZAWA-12-08\\
March, 2012
\end{flushright}
\vspace*{0.5cm}

\begin{center}

{\Large \bf 
Super Flavorsymmetry \\
with\\
Multiple Higgs Doublets\footnote{submitted to {\bf Fortschritte der Physik}
(April 2012)}}

\end{center}

\vspace{0.2cm}

\begin{center}
{\large 
Jisuke Kubo\footnote{e-mail:~jik@hep.s.kanazawa-u.ac.jp}}
\vspace {0.5cm}\\
{\it Institute for Theoretical Physics, Kanazawa University,\\
        Kanazawa 920-1192, Japan}
\end{center}
\vspace{1cm}
\begin{center}
{\Large\bf Abstract}
\end{center}
\vspace{1cm}

The meaning of non-abelian super flavorsymmetry in the presence of
Multiple Higgs Doublets is
twofold: The first one is to order the Yukawa sector of the standard model, while generating
the Cheng-Sher suppression of tree-level Flavor Changing Neutral Currents (FCNCs), and the second
one is to soften the fine tuning problem in the scalar potential and at the same time
to suppress  FCNCs coming from the soft breaking of supersymmetry.
If spontaneous CP violation is combined with non-abelian super flavorsymmetry
the CP phases in the soft supersymmetry breaking sector can disappear.
This feature of super flavorsymmetry is 
reviewed and elaborated in a supersymmetric model of flavor based 
on the finite group $Q_6$.
The predictions of the model in ferimion mass and mixing,
and flavor and CP violation can be tested in future experiments.
\vspace{0.5cm}
\newpage
\setcounter{footnote}{0}
\def\thefootnote{\arabic{footnote}}
\section{Introduction}

Introducing  flavor symmetry based on a  non-abelian finite
group into the
standard model (SM)  means going beyond
the SM \cite{Altarelli:2010gt}. 
The main reason of  the introduction of   flavor symmetry
lies in  the Yukawa sector of the SM,
because the most of the free parameters of the SM
are involved in this sector, and there exists 
within  the SM no principle how to fix its structure.
There are many guidelines to go
beyond the SM. Therefore, depending on the guideline, 
how to introduce  non-abelian flavor symmetry
 is different  \cite{Pakvasa:1977in}-\cite{Frampton:2000mq}.
The difference of the guidelines may be summarized as:
\begin{enumerate}
\item
How  it is broken.\\
Since there exists no  exact non-abelian flavor symmetry 
in the SM, it hast to  be broken:
It is spontaneously, softly or hardly broken.
\item
The scale of its breaking.\\
It is a low energy scale or a much higher scale, e.g., a GUT scale.

\item
Renormalizability.\\
The extended model is renormalizable or non-renormalizable.

\item
On which sector it
is effectively acting:
On the quark sector;  on the lepton sector;
or on both the quark  and the  lepton sector.

\end{enumerate}
\noindent
In addition to the above four categories
we can add whether the model is supersymmetric or not, 
it is a four dimensional model or not, and whether
CP is spontaneously, softly or explicitly broken.
 Note that no gauge group and no flavor group are fixed yet.
However, except for the existence of dark matter and the neutrino
mixing and masses, despite the great efforts of experimentalists,
there is  no experimental hint how to go beyond the SM.

Under these circumstances I prefer to
attach importance to the experimental verifiability of  a model.
It is clear that the lower the scale of the breaking of  flavor symmetry is,
the more verifiable is the model.
Moreover, if  flavor symmetry is hardly broken, it is not testable
in a strict sense.
Further, to decide whether the model should be  renormalizable or not, 
I would like to follow the attempt of 
Kobayashi and Maskawa \cite{Kobayashi:1973fv}, who introduced CP violation within
the {\em renormalizable extension} of the two generation SM.
 From these considerations I decided to restrict myself,  in this contribution
to the special issue, 
to  renormalizable models of flavor, where 
 non-abelian discrete flavor symmetries are
at most softly broken at low energy.
One immediately realizes that it is impossible to
present a realistic model along this line, unless one introduces
multiple Higgs doublets which belong to  non-trivial
representations of the corresponding non-abelian finite group.
Therefore,   flavor symmetry is 
at least spontaneously broken at the electroweak scale
together with the electroweak gauge symmetry.
This can lead to   severe problems
with Flavor Changing Neutral Currents (FCNCs), if
the extra Higgs bosons are not  heavy enough
\cite{Bjorken:1977vt}.
I will not discuss the details of  the different constraints
such as $\mu \to 3 e$ coming
from flavor violations  (see \cite{Atwood:1996vj} for instance). 
I simply assume that  their masses are larger than several TeV
and the electroweak precision constraints are 
satisfied 
\cite{Grimus:2007if}.

In contrast to  two Higgs doublet models
without non-abelian flavor symmetry\cite{Branco:2011iw},
the size of the Yukawa couplings are not
arbitrary, because  fermion masse and
mixing would not come out right.
However, the scale is not $O(100)$ TeV \cite{Bjorken:1977vt},
 because  the Cheng-Sher mechanism \cite{Cheng:1987rs},
 which suppresses the tree-level FCNCs,  is partially working thanks to flavor symmetry
 and making the scale of flavor symmetry decrease down to tera scale.
The presence of  extra heavy Higgs doublet bosons of tera scale
causes another problem, a fine tuning in the scalar potential.
The fine tuning can be made  stable against radiative corrections
by supersymmetry \cite{Wess:1973kz,Iliopoulos:1974zv,Nilles:1983ge}.
This is why I will concentrate on  supersymmetric models 
 of flavor.
Due to the very nature of non-renormalization theorem,
extended to  softly broken supersymmetric theories 
\cite{Ferrara:1979wa,Jack:1997pa},
the soft breaking terms of flavor symmetry
can be classified, in addition to according
to their canonical dimensions (Symanzik's theorem \cite{symanzik}).
I will discuss this matter in a general manner in Sect. 2 and 3.

The meaning of non-abelian flavor symmetry in supersymmetric models is
twofold: The first one is to introduce a principle in the Yukawa sector
as it is the case for non-supersymmetric models, while generating
the Cheng-Sher suppression, and the second
one is to  suppress FCNCs coming from
the soft breaking of supersymmetry \cite{Ellis:1981ts}-\cite{Misiak:1997ei},
while  supersymmetry softens the fine tuning problem in the scalar potential.
This mechanism to suppress FCNCs 
is  alternative to hidden sector supersymmetry breaking.
(The both mechanisms  can be of course combined.)
The suppression mechanism
has been observed in many concrete models 
\cite{Babu:2004tn,Kajiyama:2005rk,Dine:1993np}-\cite{Babu:2011mv}, and it
is in fact very general.
In certain models, like the one \cite{Babu:2004tn} which I will discuss in detail,
the  CP violating phases in the soft supersymmetry breaking (SSB)
sector  disappear,
if CP is spontaneously broken:  A phase self-alignment works \cite{Babu:2004tn}.

From Sect. 4 on I will discuss  the specific
supersymmetric flavor model based on the
finite flavor group $Q_6$
 \cite{Babu:2004tn,Kajiyama:2005rk},
\cite{Kifune:2007fj}-
\cite{Babu:2011mv}.
This model will illustrate the basic features
of any realistic supersymmetric flavor models
with multiple Higgs doublet supermultiplets \cite{Djouadi:2005gj}.
The predictions on  fermion mass and mixing
are discussed in Sect. 5, and I will show  in Sect. 6
how FCNCs and CP violations coming from the supersymmetry breaking
are suppressed by  $Q_6$ flavor symmetry.
Sect. 7 is devoted to Summary.

\section{
Flavor symmetry
and non-renormalization theorem in softly broken
supersymmetric theories}

Favor symmetry can control the structure of
the independent parameters of a theory.
In  supersymmetric theories, moreover,
the non-renormalization
 theorem allows to drop
certain  couplings
and also to  relate them with each other, 
without facing contradictions with renormalizability.
We would like to discuss 
the role of the non-renormalization
 theorem for  flavor symmetry and vice versa.
 We begin by  considering 
a generic $N=1$
supersymmetric gauge theory whose superpotential
is given by
\be
W(\Phi) &= &\frac{1}{6} Y^{ijk} \Phi_i \Phi_j \Phi_k
+\frac{1}{2} \mu^{ij} \Phi_i\Phi_j ~.
\ee
The  soft supersymmetry breaking (SSB)  Lagrangian is given by
\be
L(\Phi,W) &=& - \left( ~\int d^2\theta\eta (  \frac{1}{6} 
 h^{ijk} \Phi_i \Phi_j \Phi_k +  \frac{1}{2}  b^{ij} \Phi_i \Phi_j 
+  \frac{1}{2}  M_gW_A^\alpha W_{A\alpha} )+
\mbox{h.c.}~\right)\nn\\
& &-\int d^4\theta\tilde{\eta} \eta \overline{\Phi^j}                   
(m^2)^i_j(e^{2gV})_i^k \Phi_k~,
\ee
where $\eta = \theta^2$, 
$\tilde{\eta} = \tilde{\theta}^2$ are the external
spurion superfields and $M_g$ is the gaugino mass.

To proceed  we take into account the presence 
of  flavor symmetry.
We  recall that the D-terms are
renormalized and the wave function renormalization
can mix matter superfields $\Phi_i$ in general.
If a non-diagonal (infinite) kinetic term is induced,
a corresponding non-diagonal counter term should be added.
In what follows we assume that
the kinetic mixing among $\Phi_i$'s 
is forbidden by   flavor symmetry.
This implies that the anomalous dimensions $\gamma^i{}_j$
are diagonal, i.e.,
$\gamma^i{}_j=\delta^i{}_j~\gamma_j$.
Then the $\beta$ functions of  $Y, \mu, h$ and $m^2$
 are given by \cite{Jack:1997pa,Kobayashi:1998jq}
 \be
\beta_Y^{ijk}&=&Y^{ijk}(\gamma_i+\gamma_j+\gamma_k)~,~
\beta_\mu^{ij}=\mu^{ij}(\gamma_i+\gamma_j)~,
\label{betamu3}\\
\beta_h^{ijk}&=&(h^{ijk}-2Y^{ijk}{\cal O}) 
(\gamma_i+\gamma_j+\gamma_k)~,~
\beta_b^{ij}=(b^{ij}-2\mu^{ij}{\cal O})(\gamma_i+\gamma_j)~,\\
\label{betab3}
(\beta_{m^2})_l &=&\left[ \Delta 
+ X \frac{\partial}{\partial g}\right]\gamma_l~,~
{\cal O} =\left(M_g g^2{\partial\over{\partial g^2}}
-h^{lmn}{\partial
\over{\partial Y^{lmn}}}\right)~,
\label{diffo}\\
\Delta &=& 2{\cal O}{\cal O}^* +2|M_g|^2 g^2{\partial
\over{\partial g^2}} +\tilde{Y}_{lmn}
{\partial\over{\partial
Y_{lmn}}} +\tilde{Y}^{lmn}{\partial\over{\partial Y^{lmn}}}~,
\label{delta}
\ee
where 
 $(\gamma_1)_i={\cal O}\gamma_i$, 
$Y_{lmn} = (Y^{lmn})^*$, and $
\tilde{Y}^{ijk}=
Y^{ijk}(m^2_i+m^2_j+m^2_k)~,~
X =
\frac{-|M_g|^2 C(G)+\sum_l m_l^2 T(R_l) }{C(G)-8\pi^2/g^2}$, and
$T(R_l)$ is the Dynkin index of $R_l$, and $C_{2}(G)$
is the quadratic Casimir of the adjoint representation of the
gauge group $G$.
From these equations we may conclude
(which is basically 
Symanzik's theorem \cite{symanzik} applied to  softly broken
supersymmetric gauge theories):

\noindent
{\bf Th1}. 
The $\mu$ sector can have a 
flavor symmetry which is different from
the flavor symmetry of the Yukawa sector
if both symmetries are compatible 
with respect to renormalization
of $\mu^{ij}$. That is, if 
$\mu^{ij}(\gamma_i+\gamma_j) \sim \mu^{ij}$.

\noindent
{\bf Th2}. 
It is consistent to
introduce into the soft tri-linear couplings the same flavor symmetry
as that of the Yukawa couplings, even if it
is violated in other sectors.

\noindent
{\bf Th3}. 
The flavor symmetry which protects the kinetic mixing among $\Phi_i$'s
ensures that $(m^2)^i_j$ are diagonal.
If  the Yukawa couplings
and  soft tri-linear couplings have the
flavor symmetry,
the soft scalar mass terms, too,  can have
the flavor symmetry,
even if the $\mu$ and $b$ terms
do not respect the flavor symmetry.

\noindent
{\bf Th4}. 
The $b$ terms associated with the $\mu$ terms
should  always exist (see (\ref{betab3})).
But the $b$ sector
has no influence on the infinite renormalization
of the parameters in other sectors.
So the violation of a  symmetry in the $b$ sector 
is doubly soft.
Similarly, the soft scalar mass terms and 
 soft tri-linear couplings have no influence on the infinite
 renormalization of the $b$ terms.

\section{Softly broken flavor symmetries
and supersymmetry}
We restrict ourselves to the gauge group $SU(3)_C\times SU(2)_L\times
U(1)_Y$ and require renormalizability as it is announced.
As in the non-supersymmetric case, 
realistic softly broken supersymmetric 
models with a low energy flavor symmetry
will contain  certain number of  
Higgs doublet supermultiplets pairs $\Phi_I^u$ and $\Phi_I^d$.
The superpotential in the matter sector
is given by $W_Q+W_L+W_M$, where
\be
W_Q &=&
Y_{ij}^{uI} Q_{i} U_{j}^c  \Phi^u_I+
Y_{ij}^{dI} Q_{i} D_{j}^c   \Phi^d_I~,~
W_L
=
Y_{ij}^{eI} L_{i} E_{j}^c  \Phi^d_I+
Y_{ij}^{\nu I} L_{i} N_{j}^c   \Phi^u_I~,
\label{superP}\\
W_M&=&\frac{1}{2}M_{ij}~N_{i}^c  N_{j}^c ~,
\label{WM}
\ee
where  $Q $ and $ L$ stand for $SU(2)_L$ doublets
of the quark and lepton  supermultiplets,
respectively, where $i$ and $j$ are flavor group indices.
Similarly, $U^c, E^c$ and  $N^c$ stand for $SU(2)_L$ singlets 
of the quark, charged lepton and right-handed neutrino 
supermultiplets.
In the following discussions we denote the scalar and fermionic components
of the supermultiplets as
\be
Q &=&(~\tilde{q}_L~,~q_L~)= \left(\begin{array}{cc}
\tilde{u}_L~,& u_L\\
\tilde{d}_L~,& d_L\\
\end{array}\right)~,~U^c=(\tilde{u}_R^*~,~ u_R^c)~,
\label{superpartner1}
\ee
and similarly for $D^c, L$, etc.
The Higgs supermultiplets  are 
\be
\Phi^u &=&(~h^u~,~\tilde{h}^u)= \left(\begin{array}{cc}
h^{u+}~, &\tilde{h}^{u+}\\
h^{u0}~, &\tilde{h}^{u0}\\
\end{array}\right)~,~\Phi^d =(~h^d~,~\tilde{h}^d) = \left(\begin{array}{cc}
h^{d0}~, &\tilde{h}^{d0}\\
h^{d-}~, &\tilde{h}^{d-}\\
\end{array}\right)~,
\label{superpartner2}
\ee
and we assume that the neutral components 
of the Higgs fields, $h^{u0}$ and $h^{d0}$, acquire a 
(complex) VEV.

\noindent
\underline{The $\mu$ terms:}
The most general, renormalizable, Higgs superpotential $W_H$
has the form
\be
W_H
&=&\sum_{I,J} \mu_{IJ} \Phi^u_I \Phi^d_J~.
\label{Wmu1}
\ee
The superpotential $W_H$  often has  a symmetry larger, quite often continuos symmetry,
 than the one in the Yukawa sector.
The large continues symmetry has to be explicitly broken in the scalar potential
to avoid the appearance of the massless NG bosons and fermions.
Another problem is that 
certain fermionic components of
  $\Phi_I$ are massless. The most general form of the mass matrix ${\bf M}_N^F$ 
of the neutralinos
in the prepense of  $N$ pairs of the Higgs doublet supermultiplets can be written as
{\small   \begin{eqnarray}
{\bf M}_N^F &=&\left( \begin{array}{ccccc}
{\bf M}_{gg} & {\bf M}_{g1} & {\bf M}_{g2}  & {\bf M}_{g3} &\cdots  {\bf M}_{gN}   \\
{\bf M}_{g1}^T & {\bf M}_{11} & {\bf M}_{12} & {\bf M}_{13} &\cdots  {\bf M}_{1N}   \\
{\bf M}_{g2}^T&{\bf M}_{12}^T & {\bf M}_{22}  &   {\bf M}_{23} &\cdots  {\bf M}_{2N} \\
{\bf M}_{g3}^T &  {\bf M}_{13}^T&  {\bf M}_{23}^T
 & {\bf M}_{33} &\cdots  {\bf M}_{3N}  \\
.. & ..& .. & .. &. \\
{\bf M}_{gN}^T &\cdots&\cdots &\cdots&\cdots  {\bf M}_{NN}  \\
\end{array}\right)~,
\label{MN}
\end{eqnarray}}
where
\begin{equation}
{\bf M}_{gg} \!=\!\left( \begin{array}{cc}M_1 & \!\!0 \\
\!\!0 & M_2 \\
\end{array}\right),
{\bf M}_{gI}\! =\!
\left( \begin{array}{cc}s_W (v_{I}^u/v)M_Z &-s_W (v_{I}^d/v)M_Z  \\
-c_W (v_{I}^u/v)M_Z & c_W (v_{I}^u/v)M_Z  \\
\end{array}\right),
{\bf M}_{IJ} \!=\!\left( \begin{array}{cc}0 & \!\!-\mu_{IJ} \\
\!\!-\mu_{IJ} &0 \\
\end{array}\right),
\label{Mgg}
\end{equation}
$s(c)_W=\sin(\cos)\theta_W$, and
$M_{1,2}$ are gaugino masses.
Similarly, the mass matrix of the charginos 
${\bf M}_C$ has the form
{\small  \begin{eqnarray}
{\bf M}_C^F &=&\left( \begin{array}{cccc}
M_2 & \ldots & (v_J^d/v)M_W & \ldots \\
\vdots &   & \vdots &  \\
(v_I^u/v)M_W & \ldots & \mu_{IJ} & \ldots \\
\vdots &   & \vdots &   \\
\end{array}\right)~.
\label{MC1}
\end{eqnarray}}
To give the massless fermions a mass  we can
break 
flavor symmetry   at the superpotential level, which however should be
consistent  with {\bf Th1}.

\noindent
\underline{The soft scalar mass terms:}
Since  flavor symmetry 
for the soft scalar mass terms of the matter fields  
plays a crucial role to suppress
FCNCs in the SSB sector,
we assume that flavor symmetry is not broken 
by the soft scalar mass terms.
This requires implicitly that 
 flavor symmetry is respected by the soft tri-linear couplings,
because the  breaking by the soft tri-linear couplings
 is not consistent with the assumption that
 flavor symmetry is respected by the soft scalar mass terms.
The most general form of the soft scalar mass terms for the
multi Higgs doublet fields is
\begin{equation}
{\cal L}_{m_H^2}\!\!
=\!\!\!\sum_{I}\left( m_{H^u_I}^2 (h_I^u)^* h_I^u+
m_{H^d_I}^2 (h_I^d)^* h_I^d+
\right)\!\!
+\!\!\sum_{I<J}
\left( m_{H^u_{IJ}}^2 (h_I^u)^* h_J^u
+ m_{H^d_{IJ}}^2 (h_I^d)^* h_J^d+h.c.\right)~.
\label{m1}
\end{equation}
The superpotential (\ref{superP}) gives rise to quadratic terms, too:
\be
{\cal L}_{\mu^2}
&=&\sum_{I,J,K}~\left(\mu_{IK}^* \mu_{JK}(h_I^u)^* h_J^u+
\mu_{KI}^* \mu_{KJ}(h_I^d)^* h_J^d
\right)~.
\ee

\noindent
\underline{The $b$ terms:}\\
The $b$ sector should contain
at least terms which correspond to the $\mu$
terms in $W_H$:
\be
{\cal L}_b
&=&\sum_{IJ}~\left( b_{IJ} h_I^u h_J^d+h.c.\right)~.
\label{b1}
\ee
If  flavor symmetry should be exact, then 
$b_{IJ}$ in  (\ref{Wmu1}) have the same structure as $\mu_{IJ}$  in  (\ref{Wmu1}).
As we discussed, the breaking of  flavor symmetry
by the $b$ terms is doubly soft; it is soft, because the  canonical dimension
of the breaking operators is two, and the breaking 
has no influence on the infinite renormalization
of the soft scalar masses (see {\bf Th2}).
Therefore, if the Higgs superpotential $W_H$ given in (\ref{Wmu1}) has an  accidental continuous
 symmetry, one can break it softly by the $b$ terms
 to avoid the appearance of the NG bosons and fermions.

If the $b$ parameters are complex,
CP is explicitly broken by the $b$ terms;
the breaking is doubly soft.
CP breaking by the $b$  terms is an economic way,
because one can overcome the domain wall problem
which might appear when discrete flavor 
symmetries are spontaneously broken,
and one may moreover obtain extra CP phases to make 
electroweak baryogenesis possible.

\noindent
\underline{Softly broken CP invariance:}
In the presence of   Higgs doublet pairs only,
the quartic terms in the Higgs scalar potential come only from
the D-terms, which are  positive semi definite.
Therefore, to show that the potential minimum 
is not at the origin, one needs to consider only the quadratic
terms
\be
{\cal L}_{m^2_H}+{\cal L}_b &=& \frac{1}{2}\sum_{I,J}
\left(\begin{array}{c}
\varphi_I^u\\
\chi_I^u\\
\varphi_I^d\\
\chi_I^d\end{array}\right)^T
({\bf M_H}^2)_{IJ}\left(\begin{array}{c}
\varphi_J^u\\
\chi_J^u\\
\varphi_J^d\\
\chi_J^d\end{array}\right)~,
\label{LHb}
\ee
where $h_I^{u,d}=(\varphi_I^{u,d}+i \chi_I^{u,d})/\sqrt{2}$, and
the $4\times 4$ matrix $({\bf M_H^2})_{IJ}$ has the form
{\small  \begin{equation}
({\bf M_H^2})_{IJ} \!\!=\!\!
\left(\begin{array}{cccc}
m_{H_{IJ}^u}^2 +\mu_{IK}^*\mu_{JK}& m_{H_{IJ}^u}^2+\mu_{IK}^*\mu_{JK}
& \mbox{Re}(b_{IJ}) & -\mbox{Im}(b_{IJ})\\
m_{H_{IJ}^u}^2 +\mu_{IK}^*\mu_{JK}& m_{H_{IJ}^u}^2+\mu_{IK}^*\mu_{JK}
& -\mbox{Im}(b_{IJ}) &-\mbox{Re}(b_{IJ})\\
\mbox{Re}(b_{JI}) & -\mbox{Im}(b_{JI})& m_{H_{IJ}^d}^2 +\mu_{KI}^*\mu_{KJ} 
&m_{H_{IJ}^d}^2 +\mu_{KI}^*\mu_{KJ}  \\
-\mbox{Im}(b_{JI}) & -\mbox{Re}(b_{JI})&m_{H_{IJ}^d}^2  +\mu_{KI}^*\mu_{KJ} 
 &m_{H_{IJ}^d}^2  +\mu_{KI}^*\mu_{KJ} \\
\end{array}
\right).
\label{MHIJ}
\end{equation}}
The relevant matrix is the $4N\times 4N$ matrix ${\cal M}_H^2$
($N$ is the number of the Higgs doublet pairs)
\be
{\cal M}_H^2&=&\left(\begin{array}{ccc}
({\bf M_H^2})_{11} &({\bf M_H^2})_{11}   & \ldots \\
({\bf M_H^2})_{21}   &({\bf M_H^2})_{22}  & \ldots \\
\ldots  &\ldots  & \ldots \\
\end{array}\right)~,
\label{calM}
\ee
whose eigenvalues control the behavior of the 
Higgs scalar potential in the $4N$ dimensional space.
We find that all the eigenvalues of  ${\cal M}_H^2$ are doubly degenerate, 
 and that two orthogonal eigenvectors
 of the same eigenvalue can be always written in the form
$ {\vec u}_A = (~u_1, u_2,u_3,u_4,\ldots~)
 ~\mbox{and}~
 {\vec u}_B = (~u_2, -u_1,-u_4, u_3, \ldots~)$.
 This is due to the $U(1)_Y$ gauge invariance:
 All the directions defined by a linear combination
of  $ {\vec u}_A$ 
and $ {\vec u}_B$ are physically equivalent.
That is, one can make an appropriate linear combination
such that at least one imaginary part vanishes.
If on the other hand all the imaginary parts of $m^2_H$'s and $b$'s vanish, then 
the eigenvectors can always be  written
as 
$(~c_\theta u_1, -s_\theta u_1,
c_\theta u_3,~s_\theta u_3,c_\theta u_5,
-s_\theta u_5,\ldots~)$,
where $c_\theta=\cos\theta$ and $s_\theta=\sin\theta$,
and the angle $\theta$ is, thanks to $U(1)_Y$,  a free-choosable parameter.
Therefore,  we can always choose $\theta$ to be zero, i.e., all
the imaginary parts vanish in this direction, 
implying that
 spontaneous CP violation
at the tree level is not possible \cite{Romao:1986jy}.
 Therefore,  at least some of $m^2_H$ or/and of $b$ should be complex so that
the Higgs doublet fields  can acquire imaginary VEV. 
Needless to say that negative eigenvalues of (\ref{calM}) have to exist
for the potential minimum to
be located other than at the origin.

\noindent
\underline{The tri-linear couplings :}
The soft tri-linear couplings corresponding to the 
superpotential (\ref{superP}) is
\be
{\cal L}_A &=&
h_{ij}^{uI} \tilde{q}_{Li} \tilde{u}_{Rj}^*  h^u_I+
h_{ij}^{dI} \tilde{q}_{Li} \tilde{u}_{Rj}^*   h^d_I+
h_{ij}^{eI}\tilde{ l}_{Li} \tilde{e}_{Rj}^*  h^d_I+
h_{ij}^{\nu I} \tilde{l}_{Li} \tilde{\nu}_{Rj}^*   h^u_I+h.c.~.
\label{LA}
\ee
According to {\bf Th2} and {\bf Th3},  the soft tri-linear couplings and
soft scalar mass terms can have the same flavor symmetry 
as that of the Yukawa sector, even if flavor symmetry is softly
broken in the $\mu$ sector and by the $b$ terms.
Further, the superpotential (\ref{superP}) gives rise to tri-linear couplings, too:
\begin{equation}
{\cal L}_{\mu} =\mu_{IJ}(Y^{dJ}_{ij})^* 
\tilde{Q}_i^{\dag} \tilde{d}_{Rj}~h_I^{u}
-\mu_{IJ}(Y^{uI}_{ij})^* 
\tilde{Q}_i^{\dag} \tilde{u}_{Rj}~h_J^{d}
+\mu_{IJ}(Y^{eJ}_{ij})^* \tilde{L}_i^{\dag} \tilde{e}_{Rj}~h_I^{u}
-\mu_{IJ}(Y^{\nu I}_{ij})^* 
\tilde{L}_i^{\dag} \tilde{n}_{Rj}~h_J^{d}+h.c.~,
\label{Lmu}
\end{equation}
which contributes to the soft left-right mass matrices of the sfermions
as we will see later on.

\section{A concrete Model of flavor based on the finite group $Q_6$ \cite{Babu:2004tn}}
In Table 1  the $Q_{6}$ assignment of the quark, 
lepton and Higgs chiral supermultiplets is given
\footnote{The finite group $Q_6$ was also considered 
by Frampton and Kong in \cite{Ma:1990qh}. 
 See \cite{Babu:2004tn} for the tensor product. The same model exists for $Q_{2N}$
if $N$ is odd and a multiple  of $3$. }.
In addition to $Q_6$ flavor symmetry, a flavor universal $Z_4$ symmetry
is introduced. 
Owing to this $Z_4$, even after spontaneous symmetry breaking,  unbroken 
 interchange symmetries
${\cal P}_{I,II}$ survive in the Higgs potential \cite{Babu:2011mv}.
These symmetries ${\cal P}_{I,II}$ along with $Q_6$ enable us to reduce
significantly the number of parameters in the fermion mass matrices. 
This reduction of parameters leads to a sum rule involving quark masses and mixing
parameters 
\cite{Babu:2004tn}. 
We assume that CP violation is spontaneously broken, which is perhaps more satisfying
than the usual assumption of explicit CP violation.  Nevertheless, the dominant 
source of CP violation in the quark sector is the Kobayashi-Maskawa mechanism.  
 With the spontaneous breaking of CP, the CP problem 
  that generically exists in the SSB sector 
can be softened in a rather simple way \cite{Babu:2004tn,Kajiyama:2005rk,
Ross:2004qn,Babu:2009nn,Babu:2011mv}.

{\small  \begin{table}[ht]
\caption{Particle content of the $Q_6$ model along with 
their transformation rule under 
$Q_6 \times Z_4$.}
\vspace*{-0.05in}
$$\begin{array}{|c||c|c|c|c|c|c|c|c|c|c|c|c|c|c|c|c|c|c|c|c|c|c|c|c|c|}
\hline
~& Q & L  & Q_3 & L_3 &
 U^c,D^c,N^c,E^c & U_3^c,D_3^c & N_3^c & E_3^c
 &\Phi^{u,d} & \Phi_3^{u,d} & S & S_3 & T &T_3 &U\\
\hline
Q_6 & 2 & 2' & 1' & 1 & 2' & 1'''   &1''  &1 & 2' & 1''' & 2 & 1 & 2' &1' &1\\
\hline
Z_4 &  -i & -i & -i & -i &  1 & 1  &1&1&  i & i & -1 & -1 & + &+&+\\
\hline
R &  - & -& -& - &  - & -  &-&-&  +& + & + & + & + &+&+\\ 
\hline
\end{array}$$
\end{table}}

\subsection{The  Higgs sector}
A certain set of SM singlet Higgs fields
is needed to break   $Q_6$ flavor symmetry spontaneously while avoiding 
pseudo NG bosons  \cite{Babu:2011mv}.
 The minimal such set
involves  ${\bf 2}$, ${\bf 2}'$, ${\bf 1}'$ and two ${\bf 1}$'s of $Q_6$.
The SM singlet $S$'s  mix 
the $Q_6$ doublets $\Phi^{u,d}$ with the $Q_6$ singlets
 $\Phi_3^{u,d}$.  
An accidental $O(2)$ symmetry, which 
would exist without the $Q_6$ doublet $T$
in the Higgs potential,   is violated by
the cubic coupling of $T$.  The 
$T_3$ is needed  for the Majorana mass of $N_3^c$,
and the $Q_6$ singlet $U$ is introduced to
generate  spontaneous  CP violation  and to induce 
 spontaneous symmetry breaking of $Q_6\times Z_4$ within the 
SM singlet  sector.

The most general Higgs superpotential involving the Higgs multiplets
invariant under the $Q_6 \times Z_4$ symmetry along with the matter parity
has the form
$W_{\rm Higgs} = 
W_{U}+W_{ ST}+W_{ H}$,
where
\begin{eqnarray}
W_{ U} &=& 
\mu_{U}~ U^2+\lambda ~U^3+
\left(\lambda_1 ~ S_3^2 +\lambda_2 ~T_3^2+
\lambda_3 T\cdot T \right)U,
\label{WHiggsU}\\
W_{ ST} &=& 
\mu_{S_3}~ S_3^2+\mu_{T}~T\cdot T +\mu_{T_3} ~T_3^2
+ \lambda'_3~T\cdot(T \otimes T)\nonumber\\
& &+\lambda'_1 [~-2 S_2 S_1 T_1+(S_1^2-S_2^2)T_2~]
+\lambda'_2 S\cdot  S T_3~,\\
\label{WHiggsST}
W_{H} &=&
\lambda''_1~\Phi^u_3  (\Phi^d \ast S)+
\lambda''_2~ (\Phi^u \ast S) \Phi^d_3 +
\lambda''_3~(\Phi^u \cdot \Phi^d) S_3
\label{WHiggsH}
\end{eqnarray}
with
$x \cdot y =x_1 y_1+x_2y_2  
~,~x \ast y =x_1 y_2+x_2y_1   ~,~x \star y=x_1 y_2-x_2y_1 ~,~
x\cdot(y \otimes z) = x_1(-y_1 z_1+y_2 z_2)+x_2(y_1z_2+y_2 z_1)$.
The Higgs potential contains $F$-terms derived from 
Eqs. (\ref{WHiggsU})- (\ref{WHiggsH}), 
$D$-terms associated with
$SU(2)_L \times U(1)_Y$ gauge symmetry, and the following SSB
Lagrangian \footnote{We use the same symbol for the scalar
components as the superfields for $T, S$ and $U$.}
\begin{eqnarray}
\label{Vsoft}
{\cal L}_{\rm soft} &=& m_{U}^2 |U|^2+ m_S^2(|S_1|^2+|S_2|^2)
+ m_{S_3}^2 |S_3|^2 +
m_T^2 (|T_1|^2 + |T_2|^2) + m_{T_3}^2 |T_3|^2 
 \nonumber \\
 &+& m_{h_3^u}^2 |h_3^u|^2+
m_{h_3^d}^2 |h_3^d|^2 + 
m_{h^u}^2(|h_1^u|^2+|h_2^u|^2) + m_{h^d}^2(|h_1^d|^2+|h_2^d|^2) \nonumber \\
& +&\left\{~B_{U}~ U^2+
B_{S_3}~ S_3^2 +B_T~T \cdot T+B_{T_3}~T_3^2
\right.\nonumber \\
&+& \left[A~U^2+A_1~ S_3^2 +
A_2~ T_3^2 +A_3~ (T \cdot T) \right]~U+ A'_3~T\cdot(T \otimes T)
\nonumber \\
&+& A'_1 [~-2 S_2 S_1 T_1+(S_1^2-S_2^2)T_2~]
+A'_2 S\cdot  S T_3 \nonumber\\
& +&\left. A''_1~h^u_3  (h^d \ast S)+
A''_2~ (h^u \ast S) h^d_3 
+A''_3~(h^u \cdot h^d) S_3
+h.c.\right\}~.
\label{Lsoft}
\end{eqnarray}
We assume CP invariance, which implies that all the Yukawa couplings 
and the parameters in the Higgs potential
are  real.  
The Higgs potential 
derived from  (\ref{WHiggsU}) -(\ref{WHiggsH}), and (\ref{Vsoft}) 
including  the $D$-terms
admits two interesting minima
which leave two separate discrete symmetries 
${\cal P}_I$ and ${\cal P}_{II}$ unbroken:
\be
{\cal P}_I &:& h_1^u \leftrightarrow h_2^u,~ h_1^d \leftrightarrow h_2^d,~S_1
 \leftrightarrow S_2,~
T_2 \rightarrow -T_2,\nonumber \\
& &~h_3^u \rightarrow h_3^u,~h_3^d 
\rightarrow h_3^d,~S_3 \rightarrow S_3~,T_1 \rightarrow T_1,~T_3 \rightarrow T_3,~
~U \rightarrow U~,
\label{PI}\\
{\cal P}_{II} &:& h_1^u \leftrightarrow h_2^{u *},~h_1^d \leftrightarrow h_2^{d *},~
S_1 \leftrightarrow S_2^*,~T_2 \rightarrow -T_2^*,\nonumber \\
& &h_3^u \rightarrow h_3^{u *},~
h_3^d \rightarrow h_3^{d *},~S_3 \rightarrow S_3^*,~
~T_1 \rightarrow T_1^*,~T_3 \rightarrow T_3^*,
~U \rightarrow U^*~.
\label{PII}
\ee
The VEVs of the various Higgs fields can be consistently chosen such that these
symmetries remains unbroken.
We have explicitly found local minima at which CP and $Q_6\times Z_4$ are
spontaneously broken.
We however will not present the full analysis of  the potential.
Furthermore, in the following discussions of the paper we consider only the
second case ${\cal P}_{II}$. The ${\cal P}_{I}$ invariant case works in a similar way.

The ${\cal P}_{II}$ invariance enables us to choose a ground state given by
\begin{eqnarray}
\label{VEV2}
&~& \left\langle h_1^u \right \rangle = v_1^u e^{-i\phi^u},~
\left\langle  h_2^{u} \right \rangle = v_1^u e^{i\phi^u},~
\left\langle h_1^d \right \rangle = v_1^d e^{-i\phi^d},~
\left\langle  h_2^{d} \right \rangle = v_1^d e^{i\phi^d},~\nonumber \\
& &\left\langle h_3^u\right\rangle  = v_3^u,~
\left\langle h_3^d\right\rangle  = v_3^d,~
\left\langle S_1 \right\rangle = v_S e^{-i \phi_S},~
\left\langle S_2 \right\rangle = v_S e^{i \phi_S},~
\left\langle S_3 \right\rangle = v_{S_3},~\\
& &\left\langle T_1 \right\rangle =  v_{T_1},~
 \left\langle T_2 \right\rangle = -iv_{T_2},~
  \left\langle T_3 \right\rangle = v_{T_3},~
    \left\langle U \right\rangle = v_{U},
  \nonumber 
\end{eqnarray}
where the complex phases are all explicitly displayed.  
Note that there are only three phases, $\phi_S$, $\phi_u$ and
$\phi_d$ in the VEVs, along with a purely imaginary VEV of $T_2$.
To proceed we introduce:
\begin{eqnarray}
Y_{+}^{u,d} &=& \frac{1}{\sqrt{2}}(\Phi_1^{u,d}+\Phi_2^{u,d}),~
Y_{-}^{u,d} = \frac{i}{\sqrt{2}}(\Phi_1^{u,d}-\Phi_2^{u,d}),
\label{Gs}
\end{eqnarray}
whose VEVs are given by
$\left\langle Y_{+}^{u,d} \right\rangle =\sqrt{2}v_1^{u,d} \cos \phi^{u,d}~,
\left\langle Y_{-}^{u,d} \right\rangle =\sqrt{2}v_1^{u,d} \sin \phi^{u,d}$.
Then we redefine them as
\begin{equation}
\left( \begin{array}{c}
\Phi_L^{u}\\\Phi_H^{u}\\ \Phi_-^{u}
\end{array}
\right)=
\left(
\begin{array}{ccc}
s_{\gamma^{u}} c_{ \phi^{u}} & s_{\gamma^{u}} s_{\phi^{u}}
 & c_{\gamma^{u}}\\
c_{\gamma^{u}} c_{\phi^{u}} &c_{\gamma^{u}}s_{\phi^{u}}
 & -s_{\gamma^{u}}\\
- s_{\phi^{u}} &c_{\phi^{u}} & 0\\
\end{array}
\right)\left( \begin{array}{c}
Y_+^{u}\\Y_-^{u}\\ \Phi_3^{u}
\end{array}
\right),
\label{Phis}
\end{equation}
where
\begin{eqnarray}
c_{\gamma^{u}} &=&\cos \gamma^{u} = \sqrt{2}v_3^u/v_u~,~
s_{\gamma^{u}}= \sin \gamma^{u} = 2v_1^u/v_u~,
\label{cosgamma2}\\
c_{\phi^{u}} &=& \sin\phi^{u}~,~
s_{\phi^{u}}= \sin \phi^{u}~,~
v_u=\sqrt{2(v_3^u)^2+4(v_1^u)^2},
\label{cosphi}
\end{eqnarray}
and similarly for the down sector. As we see from 
 Eqs. (\ref{Gs})- (\ref{cosgamma2}),
only $\Phi^{u}_L$ and $\Phi^{d}_L$ have a nonvanishing VEV.
According to (\ref{superpartner2}), the new  Higgs doublet supermultiplets are defined as
\begin{eqnarray}
\Phi^u_I &=&(~h^u_I~,~\tilde{h}^u_I~)=
\left( \begin{array}{cc}
h^{u+}_I & \tilde{h}^{u+}_I\\
h^{u0}_I & \tilde{h}^{u0}_I \\
\end{array}\right)~,~
\Phi^d_I =(~h^d_I~,~\tilde{h}^d_I~)=
\left( \begin{array}{cc}
h^{d0}_I & \tilde{h}^{d0}_I\\
h^{d-}_I &\tilde{h}^{d-}_I \\
\end{array}\right)
\label{doublet}
\end{eqnarray}
with $I=L,H,-$.
As we see from  Eq. (\ref{cosgamma2}),
$h^{u0}_L$ and $h^{d0}_L$ are assumed to have 
a nonvanishing VEV: $
<h^{u0,d0}_L> =
v_{u,d}/\sqrt{2}$.
The light and heavy MSSM-like Higgs scalars  are then given by
\begin{eqnarray}
\frac{1}{\sqrt{2}} (H+i A) &=& (h^{d0}_L)^*s_\beta
- (h^{u0}_L)c_\beta~,~
\frac{1}{\sqrt{2}}( v + h+i G ) =( h^{d0}_L)^*c_\beta
+ (h^{u0}_L) s_\beta~,\label{hX}\\
G^+ &=&- (h^{d-}_L)^*c_\beta+( h^{u+}_L)s_\beta~,~
H^+ =
 (h^{d-}_L)^*s_\beta+h^{u}_Lc_\beta~,\label{HP}\\
v &=&\sqrt{v_u^2+v_d^2}~(\simeq 246 \mbox{GeV}) ~,~
\tan\beta=v_u/v_d~,
\label{tanb}
\end{eqnarray}
where $c_\beta=\cos\beta$, $s_\beta=\sin\beta$,
$G$ and $G^+$ are the NG fields, and we assumed that the mass of 
$H$ and $A$ is much larger than $M_Z$.
As in the case of the MSSM, the  couplings of $\Phi^{u,d}_L$ are
flavor-diagonal, while the extra heavy neutral fields
$h_-^{u0,d0} =
(\varphi_-^{u,d}+i \chi_-^{u,d})/\sqrt{2}~,~
h^{u0,d0}_H=(\varphi_H^{u,d}+i \chi_H^{u,d})/\sqrt{2}$
can have flavor-changing couplings.

The SM like Higgs boson $h$ defined in (\ref{hX}) is the lightest Higgs boson in this model.
The tree-level upper bound of its mass has due to
the cubic superpotential  (\ref{WHiggsH}) an extra contribution
in addition to the MSSM one:
\be
m_h^2 & \lsim & M_Z^2  c_{2\beta}^2+\frac{1}{2} v^2 s_{2\beta}^2\left[~
c_{\gamma^u}^2 s_{\gamma^d}^2 (\lambda''_1)^2+
\frac{1}{2}c_{2{\gamma^u}} s_{2{\gamma^d}}
\cos(\phi^u-\phi^d)\lambda''_1\lambda''_2\right.\nn\\
& &\left.+ s_{\gamma^u}^2 [ c_{\gamma^d}^2 (\lambda''_2)^2
+s_{\gamma^d}^2 \cos(\phi^u+\phi^d)(\lambda''_3)^2]
~\right]~,
\label{mhbound}
\ee
where $c_{\gamma^u}$, etc. are defined in (\ref{cosgamma2}).
So with the radiative correction included \cite{Okada:1990vk} one can bring $m_h$
easily to $\sim 125$ GeV,  around which  an excess of events is reported 
at LHC \cite{:2012si,Chatrchyan:2012tx}.

By integrating out the SM singlet supermultiplets,
we can obtain the effective theory.
The Higgs superpotential of the effective theory 
can be written as
\begin{eqnarray}
W^{\rm eff}_{II} &=&
\mu_{11}(\Phi^u_1 \Phi^d_1+\Phi^u_2 \Phi^d_2)+
\mu_{32}\Phi^u_3 \Phi^d_2+\mu_{32}^*\Phi^u_3 \Phi^d_1+
\mu_{13}\Phi^u_1 \Phi^d_3+\mu_{13}^*\Phi^u_2 \Phi^d_3~,
\label{weff}
\ee
where $\mu_{11}=\mu_{22}$ is real. $ W^{\rm eff}_{II}$ yields a 
${\cal P}_{II}$ invariant scalar potential, but breaks
$Q_6$ flavor symmetry softly. The soft breaking is consistent with $Q_6$
according to ({\bf Th1}). In terms of the redefined Higgs superfields
(\ref{Phis}), $W^{\rm eff}_{II}$ becomes
\be
W^{\rm eff}_{II} &=&\mu_{L}~\Phi^{u}_L \Phi^{d}_L+
\mu_{LH}~\Phi^{u}_L \Phi^{d}_H
+\mu_{HL}~\Phi^{u}_H \Phi^{d}_L
+\mu_{H}~\Phi^{u}_H \Phi^{d}_H
+\mu_{-}~\Phi^{u}_- \Phi^{d}_-\nn\\
& +& \mu_{-L}\Phi^u_- \Phi^d_L+
\mu_{L-}\Phi^u_L \Phi^d_-+\mu_{-H}\Phi^u_- \Phi^d_H+
\mu_{H-}\Phi^u_H \Phi^d_-~,
\label{Weff2}
\end{eqnarray}
and similarly the SSB Lagrangian can be written as
\begin{eqnarray}
{\cal L}^{\rm eff}_{II \rm soft} &=&m_{uL}^2~|h^{u}_L |^2
+\left[~m_{uLH}^2~h^{u*}_L h^{u}_H +h.c.\right]+
m_{uH}^2~|h^{u}_H |^2+
m_{u-}^2~|h^{u}_- |^2
\nonumber\\
&+&m_{dL}^2~|h^{d}_L |^2
+\left[~m_{dLH}^2~h^{d*}_L h^{d}_H +h.c.\right]+
m_{dH}^2~|h^{u}_H |^2+
m_{d-}^2~|h^{d}_- |^2\nonumber\\
& +&\left[~
B_{L}h^{u}_L h^{d}_L 
+B_{LH}h^{u}_L h^{d}_H
+B_{HL}h^{u}_H h^{d}_L 
+B_{H}h^{u}_H h^{d}_H
+B_{-}h^{u}_- h^{d}_-
+h.c.\right]\nn\\
&+&\left[~B_{-L}h^u_- h^d_L+
B_{L-}h^u_L h^d_-+B_{-H}h^u_- h^d_H+
B_{H-}h^u_H h^d_-~h.c.~\right]~.
\label{soft2}
\end{eqnarray}
Note that the parameters 
appearing in  Eqs.~(\ref{Weff2})  and (\ref{soft2}) are all real.  This is because 
all the fields can be redefined without
a non-trivial phase rotation 
in such way that their VEVs become real.
That is, CP is restored in the Higgs sector.
Because of the restored CP invariance in the Higgs sector, the CP even and odd
fields do not mix at the tree level. Accordingly, the mass matrix 
for the neutral  CP even  Higgs bosons assumes the form
{\small \begin{eqnarray}
\textbf{M}^{2}_{N\mbox{Heven} }=
\left(
\begin{array}{cccccc}
c^{2}_{2\beta }M^{2}_{z} & -c_{2\beta }s_{2\beta }M^{2}_{z} & 0&0&0&0\\
-c_{2\beta }s_{2\beta }M^{2}_{z} &\frac{2B_{L}}{S_{2\beta}} +s^{2}_{2\beta }M^{2}_{z} & -\frac{\hat{m}^{2}_{uLH}}{c_{\beta}} &\frac{\hat{m}^{2}_{dLH}}{s_{\beta}} &
\frac{\mu _{1}^2}{c_{\beta}}&-\frac{\mu_{2}^2}{s_{\beta}}\\
0& -\frac{\hat{m}^{2}_{uLH}}{c_{\beta}} & -\hat{m}^{2}_{uH}&-B_{H}
&\mu_{3}^2&-B_{H-}\\
0 & \frac{\hat{m}^{2}_{dLH}}{s_{\beta}}& -B_{H}&-\hat{m}^{2}_{dH}&-B_{-H}&\mu_{4}^2\\
0 & \frac{\mu _{1}^2}{c_{\beta} }& \mu_{3}^2 &-B_{-H}
&-\hat{m}^{2}_{u-}&-B_{-}\\
0 & -\frac{\mu_{2}^2}{s_{\beta}} & -B_{H-} &\mu_{4}^2
&-B_{-}&-\hat{m}^{2}_{d-}\\
\end{array}
\right)
\label{MRe}
\end{eqnarray}}
in the basis
$\left( h,H,\varphi^{u}_{H},\varphi^{d}_{H},
\varphi^{u}_{-},\varphi^{d}_{-} \right)$, where
{\small  \begin{eqnarray}
\mu _{1}^2 &=&
\mu _{L-}\mu _{-}+\mu _{LH}\mu _{-H}+\mu _{L}\mu _{-L}~,~
\mu _{2}^2=
\mu _{HL}\mu _{H-}+\mu _{L}\mu _{L-}+\mu _{-}\mu _{-L}~,\\
\label{mu12}
-\hat {m}^{2}_{uLH} &=& -m^{2}_{uLH}+\mu _{HL}\mu _{L}
+\mu _{H}\mu _{LH}+\mu _{H-}\mu _{L-}~,\nonumber\\
-\hat {m}^{2}_{dLH}&=&-m^{2}_{dLH}+\mu _{H}\mu _{HL}
+\mu _{L}\mu _{L-}+\mu _{-H}\mu _{-L}~,\label{mudLH}\\
\mu _{3}^2&=&
\mu _{H-}\mu _{-}+\mu _{H}\mu _{-H}+\mu _{HL}\mu _{-L}~,~
\mu _{4}^2 = 
\mu _{H}\mu _{H-}+\mu _{LH}\mu _{L-}+\mu _{-}\mu _{-H}~,\\
-\hat {m}^{2}_{u(d)H} &=& -m^{2}_{u(d)H}+\mu^{2} _{H}+\mu^{2} _{HL(LH)}
+\mu^{2} _{H-(-H)}-(+)\frac{1}{2}c_{2\beta}M^{2}_{z}~,\nonumber\\
-\hat {m}^{2}_{u(d)-} &=& -m^{2}_{u(d)-}+\mu^{2} _{-}+\mu^{2} _{-H(H-)}
+\mu^{2} _{-L(L-)}-(+)\frac{1}{2}c_{2\beta}M^{2}_{z}~.
\label{hatmudH-}
\end{eqnarray}}
The mass matrix for the CP odd neutral Higgs bosons is found to be
{\small \begin{eqnarray}
\textbf{M}^{2}_{N\mbox{Hodd}}\!=\!\!
\left(\!\!\!
\begin{array}{ccccc}
\frac{B_{L}}{c_{\beta}s_{\beta}}&-\frac{\hat{m}^{2}_{uLH}}{c_{\beta}} &\frac{\hat{m}^{2}_{dLH}}{s_{\beta}}&\frac{\mu _{1}^2}{c_{\beta}}&
\frac{\mu_{2}^2}{s_{\beta}}\\
-\frac{\hat{m}^{2}_{uLH}}{c_{\beta}}&\!\!-\hat{m}^{2}_{uH}\!+\!c^{2}_{\theta _{W}}c_{2\beta}M^{2}_{Z}&B_{H}&\mu_{3}^2&B_{H-}\\
\frac{\hat{m}^{2}_{dLH}}{s_{\beta}}&B_{H}&\!\!-\hat{m}^{2}_{dH}\!-\!c^{2}_{\theta _{W}}c_{2\beta}M^{2}_{Z}&B_{-H}&\mu_{4}^2\\
\frac{\mu _{1}^2}{c_{\beta}}&\mu_{3}^2&B_{-H}&\!\!-\hat{m}^{2}_{u-}\!+\!c^{2}_{\theta _{W}}c_{2\beta}M^{2}_{Z}&B_{-}\\
\frac{\mu_{2}^2}{s_{\beta}}& B_{H-} & \mu_4^2 &B_{-}&\!\!\!-\hat{m}^{2}_{d-}\!\!-\!
c^{2}_{\theta _{W}}c_{2\beta}M^{2}_{Z}
\end{array}
\!\!\!\right)
\label{MIm}
\end{eqnarray}}
in the basis 
$\left( A,\chi_H^{u0},\chi_H^{d0},\chi_-^{u0},
\chi_-^{d0}\right)$.
Similarly, we obtain the mass matrix of the charged Higgs bosons:
{\small  \begin{eqnarray}
\textbf{M}^{2}_{C}=
\left(
\begin{array}{ccccc}
\frac{B_{L}}{c_{\beta}s_{\beta}}&
-\frac{\hat{m}^{2}_{uLH}}{c_{\beta}}&-\frac{\hat{m}^{2}_{dLH}}{s_{\beta}}&
\frac{\mu _{1}^2}{c_{\beta}}&\frac{\mu_{2}^2}{s_{\beta}}\\
-\frac{\hat{m}^{2}_{uLH}}{c_{\beta}}&-\hat{m}^{2}_{uH}&B_{H}&
\mu_{3}^2&B_{H-}\\
-\frac{\hat{m}^{2}_{dLH}}{s_{\beta}}&B_{H}&
-\hat{m}^{2}_{dH}&B_{-H}&\mu_{4}^2\\
\frac{\mu _{1}^2}{c_{\beta}}&\mu_{3}^2&B_{-H}&-\hat{m}^{2}_{u-}&B_{-}\\
\frac{\mu_{2}^2}{s_{\beta}}&B_{H-}&\mu_{4}^2&B_{-}&-\hat{m}^{2}_{d-}\\
\end{array}
\right)\label{MC}
\end{eqnarray}}
in the basis $\left( H^{+},h^{u+}_{H},h^{d-*}_{H},h^{u+}_{-},h^{d-*}_{-}\right)$.

\subsection{The Yukawa sector}
The most general, renormalizable,
$Q_{6} \times Z_4 \times R$ invariant superpotential $W_Y$ 
in the Yukawa sector is given by
\footnote{The $Q_{6} \times Z_4 \times R$  assignment is given in Table 1.}
$W_Y = W_Q+W_L$,
where $W_Q$ and $W_L$ are given in (\ref{superP}) 
with the Yukawa matrices
{\small 
\begin{equation}
\label{Yuq}
{\bf Y}^{u1(d1)} \!=\!\left(\!\begin{array}{ccc}
0 & 0 & 0 \\
0 & 0 & Y_b^{u(d)} \\
0&  Y_{b'}^{u(d)}  & 0 \\
\end{array}\!\!\right),~
{\bf Y}^{u2(d2)}\! =\!\left(\!\begin{array}{ccc}
0 & 0 & Y_b^{u(d)}\\
0 & 0 & 0 \\
  -Y_{b'}^{u(d)} &0 & 0 \\
\end{array}\!\!\right),\\
{\bf Y}^{u3(d3)}\!=\!\left(\!\begin{array}{ccc}
0 & Y_c^{u(d)} & 0\\
Y_c^{u(d)} & 0 & 0 \\
0 &  0 & Y_a^{u(d)} 
\end{array}\!\!\right)~,
\end{equation}
\begin{equation}
{\bf Y}^{e1} =\left(\begin{array}{ccc}
-Y_c^{e} & 0 & Y_b^{e}\\
0 & Y_c^{e} &  0\\
Y_{b'}^{e}& 0  & 0 \\
\end{array}\right),~
{\bf Y}^{e2} =\left(\begin{array}{ccc}
0 & Y_c^{e} & 0 \\
Y_c^{e} & 0 & Y_b^{e} \\
0&  Y_{b'}^e & 0 \\
\end{array}\right)~,~
{\bf Y}^{e3}=0~,
\label{Yue}
\end{equation}
\begin{equation}
\label{Yun}
{\bf Y}^{\nu1} =\left(\begin{array}{ccc}
-Y_c^{\nu}& 0 & 0 \\
0 & Y_c^{\nu} & 0 \\
 Y_{b'}^\nu & 0 & 0 \\
\end{array}\right),~
{\bf Y}^{\nu2} =\left(\begin{array}{ccc}
0 & Y_c^{\nu} & 0\\
Y_c^{\nu} &  & 0 \\
0&Y_{b'}^{\nu}  & 0 \\
\end{array}\right),
{\bf Y}^{\nu3}=\left(\begin{array}{ccc}
0 & 0 & 0\\
0 & 0 & 0 \\
0 &  0 & Y_a^{\nu} \\
\end{array}\right)~.
\end{equation}}
All the parameters appearing above are  real, because 
we assume  that   CP is spontaneously
broken.
From the Yukawa interactions along with the Yukawa matrices
(\ref{Yuq})-(\ref{Yun}) 
and the VEV structure of  the  ${\cal P}_{II}$ invariant case
(given in (\ref{VEV2}),
we obtain the fermion mass matrices.
As we will see that the resulting mass matrix of the quarks  is of 
 a nearest neighbor interaction (NNI) type
which is one of the successful Ans\" atze 
 for the quark mass matrix \cite{Fritzsch:1977vd,Branco:1988iq}
 \footnote{See \cite{Fritzsch:1999ee} for a review 
 for different Ans\" atze.}.
The $Q_{6}$ assignment of the fermions is so chosen that
the NNI type of the mass matrix occurs from  flavor symmetry
( see also \cite{Branco:2010tx} ).
  The Majorana mass term for $N^c$ is given by
$W_M = \frac{1}{2}M_1(N^c_1N^c_1+N^c_2 N^c_2  )+\lambda_M <\!\! T_3 \!\!> 
N_3^c N_3^c$.
We first discuss the quark sector and then  the lepton sector
in the following sections.

\section{Fermion mass and mixing}
\subsection{The quark sector and the Cabibbo-Kobayashi-Maskawa (CKM) mixing}

In the background ${\cal P}_{II}$,
the fermion mass matrices ${\bf m}_{u,d}$  take the form
{\small  \begin{eqnarray}
\label{Mud2}
{\bf m}_{u,d} = \left( \begin{array}{ccc} 0 & C_{u,d} & {B_{u,d} \over \sqrt{2}} 
e^{-i \phi^{u,d}} \cr C_{u,d}& 0 & {B_{u,d} \over \sqrt{2}} 
e^{i \phi^{u,d}}  \cr
-{B'_{u,d} \over \sqrt{2}}e^{-i \phi^{u,d}} & {B'_{u,d} 
\over \sqrt{2}} e^{i \phi^{u,d}} & A_{u,d}   \end{array}\right)~,
\end{eqnarray}}
where
$A_{u,d} = Y^{u,d}_a~ v_3^{u,d}~,~B_{u,d}= 
\sqrt{2}Y^{u,d}_b~ v_1^{u,d}~,~B'_{u,d} = 
\sqrt{2}Y^{u,d}_{b'} ~ v_1^{u,d}~,~C_{u,d} = Y^{u,d}_c~ v_3^{u,d}$.
The phases in the matrix of Eq. (\ref{Mud2}) can be factorized as
${\bf m}_{u,d}^r = P_{u,d} {\bf m}_{u,d} P_{u,d}$,
where
\begin{equation}
P_{u,d} = {\rm diag.}\{e^{i \phi^{u,d}},~e^{-i \phi^{u,d}},~1 \}~.
\label{Pud}
\end{equation}
Then we do a 45 degrees rotation in the (1-2) plane to bring 
${\bf m}^r_{u,d}$ into 
{\small \begin{eqnarray}
\label{Mudhat}
\hat{{\bf m}}_{u,d} =  R_L^T~{\bf m}^r_{u,d}~ R_R 
=\left(  \begin{array}{ccc}0 & C_{u,d} & 0 \cr -C_{u,d} & 0 & B_{u,d}   \cr
0 & B'_{u,d} & A_{u,d}\end{array}
 \right)~,
\end{eqnarray}}
generating  a non-trivial quark mixing matrix given by
{\small \begin{eqnarray}
K = R_L^T P_u^\dag P_d R_L=
\left(\begin{array}{ccc}\cos\phi & i \sin\phi & 
0 \cr i \sin\phi & \cos \phi & 0 \cr 0 & 0 & 1\end{array}\right)~,
~\phi=\phi^u-\phi^d~,
\label{K}
\end{eqnarray}}
where 
{\small \be
R_L &=&  \frac{1}{\sqrt{2}}\left( \begin{array}{ccc}
1 & 1 & 0\\-1 & 1& 0\\
 0 & 0 &\sqrt{2}
\end{array}\right)~,~
R_R =  \frac{1}{\sqrt{2}}\left( \begin{array}{ccc}
-1 & -1 & 0\\-1 & 1& 0\\
 0 & 0 &\sqrt{2}
\end{array}\right)~.
\label{R}
\ee}
The CKM mixing matrix is then obtained as
$V_{\rm CKM} = O_u^T K O_d$,
where $O_{u,d}$ diagonalize the matrices of Eq. (\ref{Mudhat}).
An excellent fit is obtained with the following choice of parameters at
$\mu = 1$ TeV:
\begin{eqnarray}
\label{input2}
 A_u/m_t &=&7.031\times 10^{-3},
  B_u/m_t =-3.124\times 10^{-3},B'_u/m_t = 1.000,
~C_u /m_t= 1.750\times 10^{-3},\nonumber \\
A_d/m_b &=& 0.9200,B_d /m_b= 0.04406,
~ B'_d/m_b =0.3899, C_d /m_b=4.203\times 10^{-3},\nonumber\\
\phi  &= &0.08375~.
\label{bestPII}
\end{eqnarray}
 The resulting quark masses  at $\mu = 1$ TeV are:
\begin{eqnarray}
&~& m_u = 0.910~{\rm MeV}~,~ m_c = 542~ {\rm MeV}~,~
m_d = 2.21 ~{\rm MeV}~,~ m_s =44.5 ~{\rm MeV}~,
\label{masses1new}
\end{eqnarray}
where we have used $m_t = 151.3 ~{\rm GeV}$ and
$m_b = 2.46 ~{\rm GeV}$.
These values are to be compared with quark masses extrapolated
from low energy scale to $\mu = 1$ TeV \cite{Xing:2007fb}:
\begin{eqnarray}
&&m_u=0.85 \sim 1.55\ {\rm MeV}~,~
m_d=2.05\sim 2.85\ {\rm MeV}\ ,\nonumber\\
&&m_s=39.6\sim 64.4\ {\rm MeV}~,~
m_c=502\sim 570\ {\rm MeV}\ ,\nonumber\\
&&m_b=2.39\sim 2.53\ {\rm GeV}~,~
m_t=150.3\sim 151.8\ {\rm GeV}~,
\label{masses0}
\end{eqnarray}
where we have updated the result of \cite{Xing:2007fb}
by using the updated quark masses given in PDG 2011 \cite{Nakamura:2010zzi},
while neglecting the uncertainties due to the RG running.
The input values of Eq. (\ref{input2}) give the  output for the
CKM parameters:
\begin{eqnarray}
  \lambda & =& 0.2260~,~A=0.792~,~
\bar{\rho}=0.148~,~ \bar{\eta}=0.331~,\nonumber\\
  \sin 2\beta &=& 0.675~,~
\alpha=92.8~\mbox{[deg]}~,~
  \beta=21.2~\mbox{[deg]}~,~
  \gamma=65.9\mbox{[deg]}~,
\end{eqnarray}
which should be compared with the fit result of
the CKMfitter group \cite{CKMfitter,Lenz:2012az}
\begin{eqnarray}
\lambda &=& 0.22539^{ +0.00062}_{-0.00095}~,~
A=0.801^{+0.026}_{-0.014}~,\label{lambda}\\
\bar{\rho} &=& 0.144^{+0.023}_{-0.026}~,~
\bar{\eta}=0.343^{+0.015}_{-0.014}~,~
\sin 2\beta=0.691^{+0.020}_{-0.020}~,\nonumber\\
\alpha &=& 90.9^{+3.5}_{-4.1}~\mbox{[deg]}~,~
\beta=21.84^{+0.80}_{-0.76}~\mbox{[deg]}~,~
\gamma=67.3^{+4.2}_{-3.5}~\mbox{[deg]}~.
\label{ckmfitter}
\end{eqnarray}

\begin{figure}
\includegraphics[width=9cm]{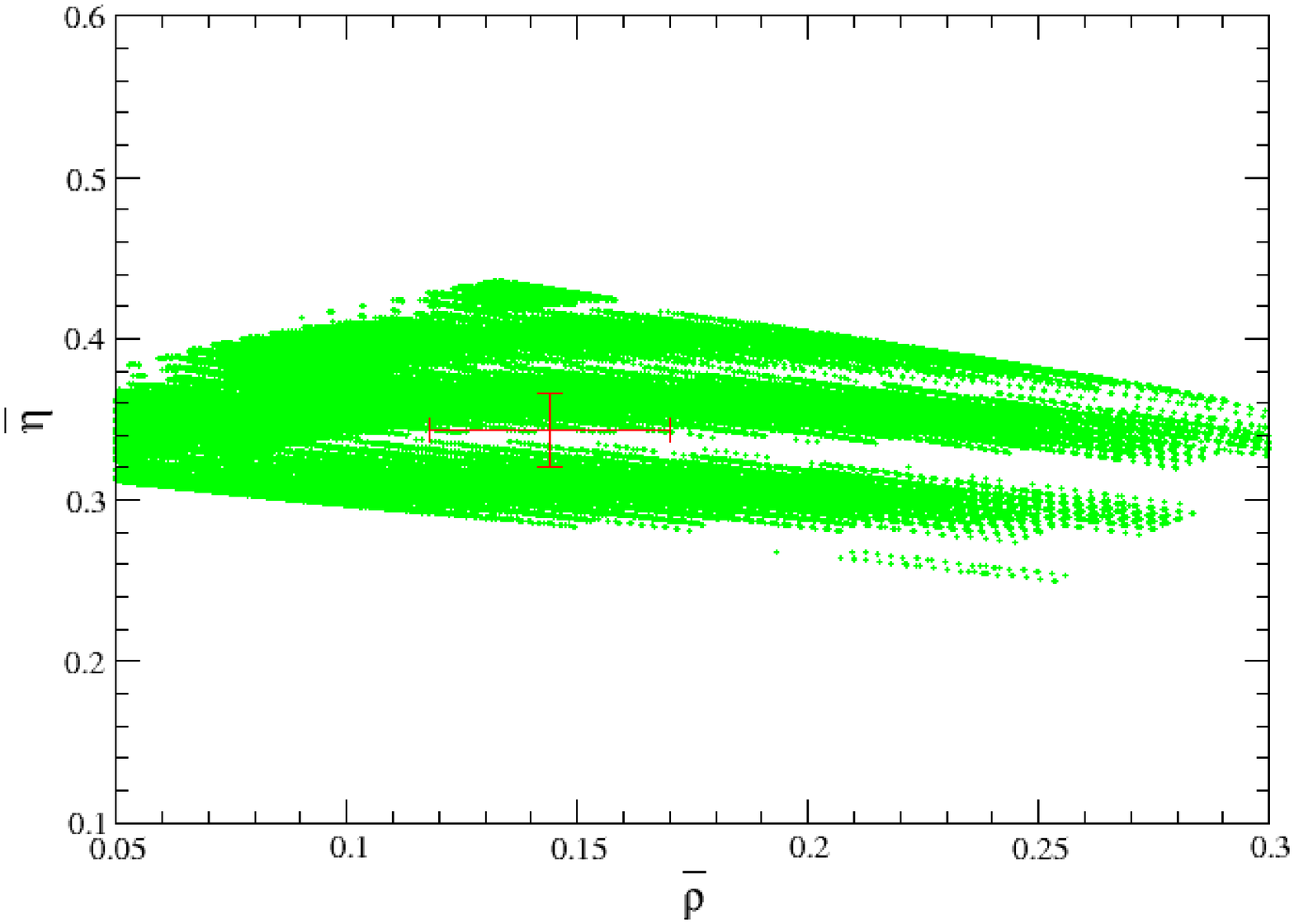}a)
\hfil
\includegraphics[width=9cm]{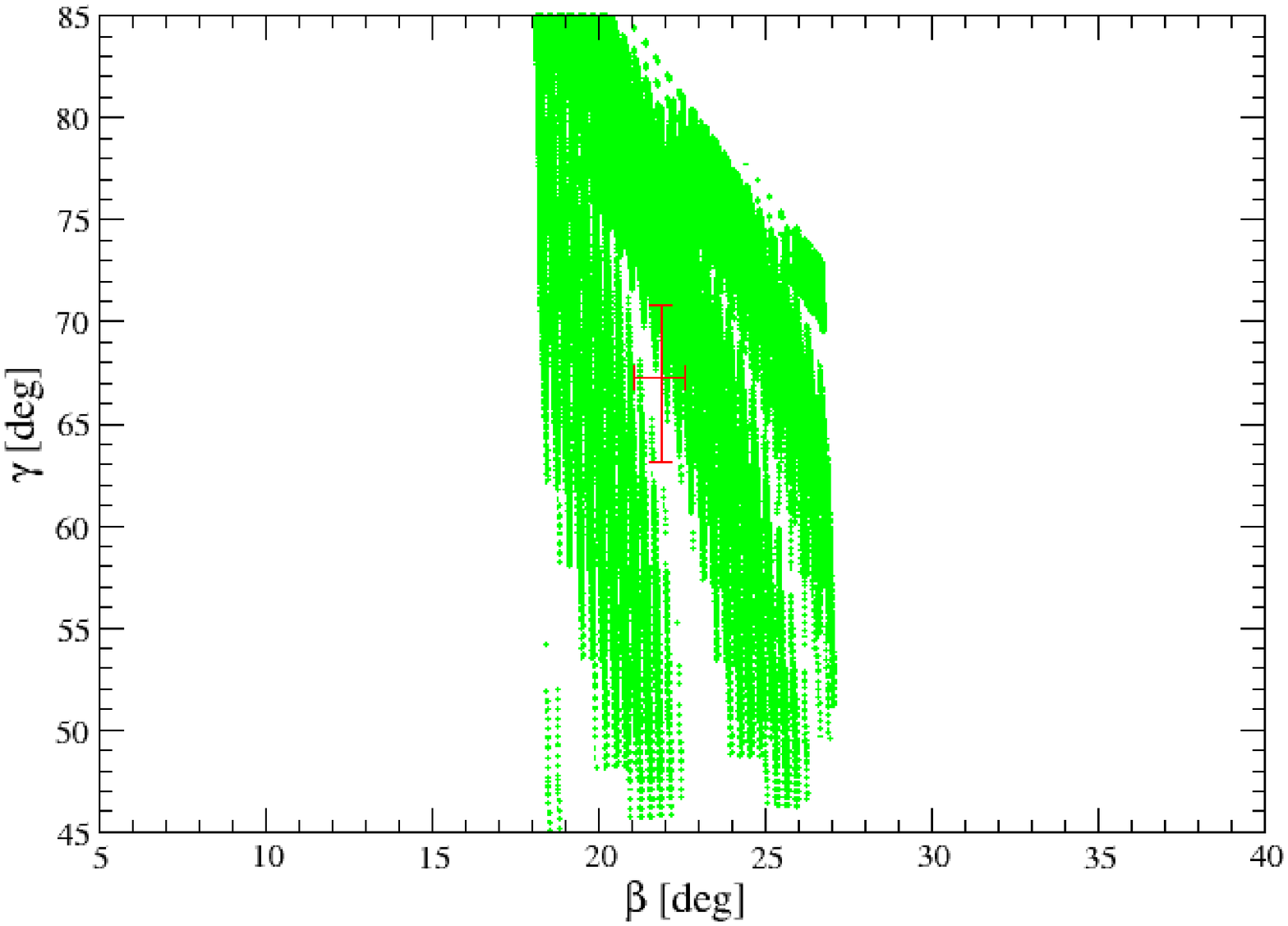}b)
\caption{ \footnotesize{The predictions in the
$\bar{\rho}-\bar{\eta}$ plane (a) and 
$\beta-\gamma$ plane (b) for the ${\cal P}_{II} $ invariant model,
where we have used
as the input parameters; the quark masses,
$\lambda$ and $A$ given in Eqs. (\ref{masses0}) and (\ref{lambda}), respectively.
We also have imposed the constraints on the quark masses \cite{Nakamura:2010zzi}:
$2m_s/(m_u+m_d)=22\sim 30~,~m_s/m_d=17\sim 22~,
m_u/m_d=0.35\sim 0.60~,~(1/2)(m_u+m_d)(2\mbox{GeV})=(3.0\sim 4.8)~\mbox{MeV}$.
The crosses denote one $\sigma$ value given in \cite{CKMfitter}.}}
\label{prediction1}
\end{figure}

Fig.~\ref{prediction1} shows the prediction in the
$\bar{\rho}-\bar{\eta}$ plane (a) and in the
$\beta-\gamma$ plane (b) for  ${\cal P}_{II}$.
The CKMfitter group best-fit value (\ref{ckmfitter})
is  indicated in these plots.
We see from  Eqs. (\ref{masses1new}) and 
Fig.~\ref{prediction1}  that
the model ${\cal P}_{II}$ reproduces the quark masses, CKM mixings and the CP violating
phase in an excellent way.

\subsection{Lepton sector}
From the VEV structure of the  ${\cal P}_{II}$ invariant case, given in (\ref{VEV2}), 
we see that the Majorana mass term for $N_3^c$ is real,
because $<\!T_3\!>$ is real:
$M_N = (M_1~,~M_1~,~ M_3 = a_{\nu^c} ~v_{T_3})$.
The Dirac neutrino and charged lepton mass matrices are:
{\small  \begin{eqnarray}
{\bf m}_{\nu^D} &=& \left( \begin{array}{ccc}-C_{\nu}e^{-i \phi^u} & C_{\nu}e^{i \phi^u} & 0 \cr
 C_{\nu}e^{i \phi^u} &  C_{\nu}e^{-i \phi^u} & 0   \cr
B'_{\nu}e^{-i \phi^u} & B'_{\nu}e^{i \phi^u} & A_{\nu}
\end{array}\right)~,~
{\bf m}_{\ell} = \left(\begin{array}{ccc}-C_{\ell}e^{i \phi^d} & C_{\ell}e^{-i \phi^d}
 & B_{\ell}e^{i \phi^d} \cr C_{\ell}e^{-i \phi^d}
& C_{\ell}e^{i \phi^d} & B_{\ell}e^{-i \phi^d}   \cr
 B'_{\ell}e^{i \phi^d} & B'_{\ell}e^{-i \phi^d} & 0
 \end{array}\right)~,
 \label{Mlep2}
\end{eqnarray}}
which (by the seesaw formula)  lead to the light neutrino Majorana mass matrix
\begin{eqnarray}
\label{Mnulight2}
M'_\nu = m_0 \left(\begin{array}{ccc} 2 \rho^2_2 \cos (2\phi^u) & 0
& 2 i \rho_2\rho_4 \sin (2\phi^u) \cr
0  & 2 \rho^2_2 \cos (2\phi^u) & 2 \rho_2\rho_4 \cr
 2 i \rho_2\rho_4 \sin (2\phi^u)  & 2 \rho_2\rho_4  
 & -\rho_3^2+2 \rho^2_4 \cos (2\phi^u)
\end{array}\right)~,
\end{eqnarray}
where
$\rho_2^2=(C_{\nu})^2 / M_1~,~
\rho_3^2 = -(A_{\nu})^2/M_3~,~\rho_4^2 =(B'_{\nu})^2/M_1$.
We have assumed that $ M_1$ is positive, while $M_3$ is negative.
\begin{figure}
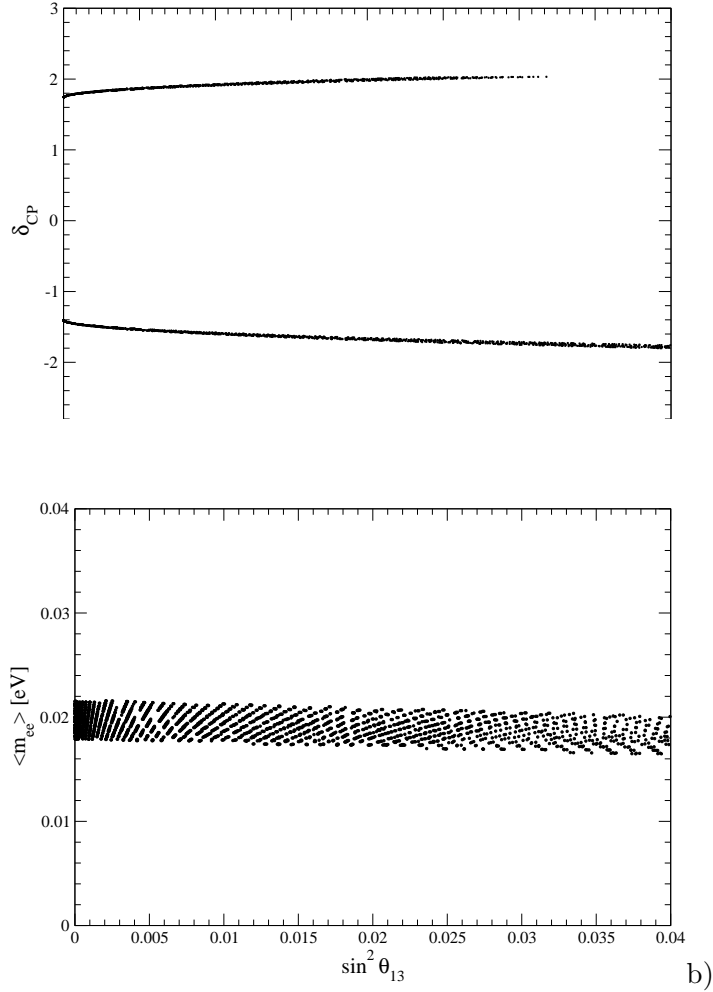

\includegraphics[width=9cm]{u13-d.eps}a)
\hfil
\includegraphics[width=9cm]{u13-mee.eps}b)
\caption{ \footnotesize{The prediction in the
$\sin^2\theta_{13}-\delta_{CP}$ plane (a) 
and in the
$\sin^2\theta_{13}-<m_{ee}>$ 
 plane  (b) for  ${\cal P}_{II}$.
We have used
 $\sin^2\theta_{12}~,~
\Delta m_{21}^2$ and 
$\Delta m_{23}^2$ given in (\ref{oscillationP})
as the input values of the parameters, and
$\phi_d=\phi_u-0.08375$.}}
\label{prediction-n-1}
\end{figure}
\begin{figure}
\includegraphics[width=9cm]{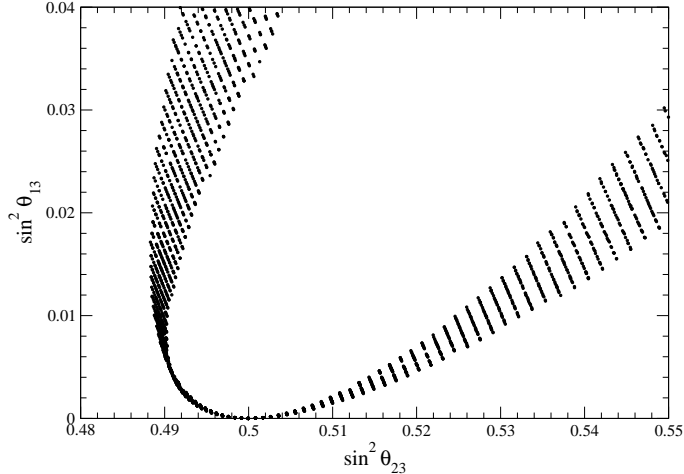}%
\caption{ \footnotesize{$\sin^2\theta_{13}$ against $\sin^2\theta_{23}$
  in the case of  ${\cal P}_{II}$
for the same input parameters as Fig.~\ref{prediction-n-1}.}}
\label{prediction-n-3}
\end{figure}
We make the matrix (\ref{Mnulight2})  real by redefining $\nu_1=i \nu_1'$.
The resulting mass matrix can
be diagonalized by an orthogonal matrix.
As for the charged lepton mass matrix ${\bf m}_{\ell}$, we can obtain  hierarchical masses, e.g.,
$m_e \sim B'_{\ell}~,~m_\mu \sim C_{\ell}~,~
m_\tau \sim B_{\ell}$. 
Since the relative phase $\phi=\phi^u-\phi^d$ is fixed in the quark sector,
there are seven independent parameters in the lepton sector.
When $\phi^u=0$, the neutrino mass matrix above  yields only a tiny
$\sin^2\theta_{13}\simeq  m_e/m_\mu \sim 10^{-3}$ \cite{Kubo:2003iw},
which  is too small 
  \cite{Schwetz:2011qt,Abe:2011sj,Adamson:2011qu,Abe:2011fz,An:2012eh}.
For non-zero $\phi^u$, we obtain $\sin\theta_{13}
\propto  \sin 2\phi^u$, which can be small
or large.
We use the three charged lepton masses, 
$\sin^2\theta_{12}~,~
\Delta m_{21}^2$ and $\Delta m_{23}^2$, and vary the value of
$\phi^u$ while $\phi^d$ is set equal to $\phi^u-0.08375$  (see (\ref{input2})),
where the up-dated  neutrino oscillation
parameters \cite{Schwetz:2011qt} in which the recent results 
of T2K \cite{Abe:2011sj} and MINOS \cite{Adamson:2011qu}
experiments are included are:
\be
\Delta m_{21}^2 &= &7.59^{+0.20}_{-0.18} \times10^{-5}~\mbox{ eV}^2~,~
\Delta m_{23}^2=2.4^{+0.08}_{-0.09} 
\times 10^{-3}\mbox{ eV}^2~,\nn\\
\sin^2\theta_{12} &= &0.312^{+0.017}_{-0.015}~, ~
\sin^2\theta_{13}=0.016^{+0.008}_{-0.006}~,~
\sin^2\theta_{23}=0.52\pm 0.06~.
\label{oscillationP}
\ee
We below display the predictions of the model in different
two dimensional planes.
Fig.~\ref{prediction-n-1} (a)  shows the Dirac phase $\delta_{CP}$ (in the convention
of Ref. \cite{Nakamura:2010zzi}) against $\sin^2\theta_{13}$.
We see that the model predicts
 nearly maximal CP violation.
It is also possible to predict the
effective neutrino mass $<\!m_{ee}\!>$
as a function of $\sin^2\theta_{13}$.
If the mixing is tri-bimaximal \cite{Harrison:2002er}, for instance, then 
$m_{\nu_2}\simeq \sqrt{\Delta m_{23}^2} 3 \sqrt{3}/2 \sqrt{2}\simeq
0.09$ eV.
In Fig.~\ref{prediction-n-1} (b) we plot the prediction in the
$\sin^2\theta_{13}-<m_{ee}>$ plane.
The deviation from the maximal mixing
has terms proportional to $m_e/m_\mu$ and to
 $\sin 2 \phi_u$. In Fig.~\ref{prediction-n-3} we plot 
 $\sin^2\theta_{13}$ against
 $\sin^2\theta_{23}$. Once the $\sin^2\theta_{13}$ and
$\sin^2\theta_{23}$ are measured, the model  prediction
can be tested by experiments
 (see  \cite{Ardellier:2006mn} and \cite{Elliott:2012sp}).

\section{Suppressing FCNCs by flavor symmetry and
CP violations by self-alignment of phases}
\subsection{Tree-level FCNCs and the Cheng-Sher mechanism}
Before we go to FCNCs coming from the supersymmetry breaking, 
we shall demonstrate that the Cheng-Sher mechanism \cite{Cheng:1987rs} to suppress
the tree-level FCNCs is partially embedded
in  the $Q_6$ model (see also  \cite{Branco:2010tx}).
Consider the mass difference of the neutral Kaons
$\Delta M_K\simeq 3.5 \times 10^{-15}$ GeV, for instance,
and assume that there is a neutral Higgs boson $H$
with mass $M_H$ which has a flavor changing Yukawa coupling
 with the down and  strange quarks:
$(Y_{ds}\bar{d}_L s_R+Y_{sd}\bar{s}_L d_R) H+h.c.$.
Then the contribution to $\Delta M_K$ can be estimated as \cite{Gabbiani:1996hi}
\begin{equation}
\Delta M_K  \! \!=\! \!2 |<\!\bar{K}^0|\bar{s}_R 
d_L \bar{s}_L d_R |K^0\!>| \left( \frac{|Y_{ds} Y_{sd}^*|}{M_H^2} \right)
 \!= \!\frac{1}{2}f_K^2 B'_K M_K \left(\frac{M_K}{m_s+m_d}\right)^2
\left( \frac{Y_{ds} Y_{sd}^*}{M_H^2} \right),
\label{DMK}
\end{equation}
where 
$f_K\simeq 0.16$ GeV is the Kaon decay constant, $B'_K$ is a bag parameter
for the matrix element in (\ref{DMK}). Using  $M_K\simeq 0.5$ GeV,
 $m_s\simeq 0.1$ GeV and $m_d\simeq 5 \times 10^{-3}$ GeV, we 
 obtain
 \be
 \Delta M_K &\simeq & 
 0.56 \left( \frac{|Y_{ds} Y_{sd}^*|}{M_H^2} \right)~\mbox{GeV}~,
 \label{DMK1}
\ee
where we have assumed that $B'_K=1$.
 From (\ref{DMK1}) we obtain a lower bound for $M_H$:
$ M_H \gsim  1.3\times 10^{4}
\sqrt{|Y_{ds} Y_{sd}^*|}~~\mbox{TeV}$.
Therefore, if $|Y_{ds}|, |Y_{sd}|\simeq \sqrt{m_s m_d}/v\sim 10^{-4}$ 
( $v=246$ GeV) is satisfied,
then the lower bound reduces to $\sim O(1)$ TeV.
Cheng and Sher  showed that this is true in more general cases:
If the Yukawa couplings  satisfy
$({\bf Y})_{ij} \sim \sqrt{m_i m_j} /v$,
then the lower bound of the mass of  flavor changing neutral Higgs bosons
is  $O(\mbox{few})$ TeV.

We will show that in the $Q_6$ model this mechanism
is partially embedded.
To this end,
we consider the Yukawa matrices (\ref{Yuq}) and find:
{\small \be
{\bf Y}^{u1}&\simeq &\left(\begin{array}{ccc}
0 & 0 & 0 \\
0 & 0 & 0.05\sqrt{m_c m_t} \\
0&  17\sqrt{m_c m_t}  & 0 \\
\end{array}\right)\frac{\sqrt{2}}{v s_{\gamma^u}s_\beta} ~,~
{\bf Y}^{u2}  \simeq\left(\begin{array}{ccc}
0 & 0 & 1.3\sqrt{m_u m_t}\\
0 & 0 & 0 \\
 -408\sqrt{m_u m_t} &0 & 0 \\
\end{array}\right)\frac{\sqrt{2}}{v s_{\gamma^u}s_\beta}~,\nn\\
{\bf Y}^{u3}&\simeq&\left(\begin{array}{ccc}
0 & 12\sqrt{m_u m_c} & 0\\
12\sqrt{m_u m_c} & 0 & 0 \\
0 &  0 & 0.007\sqrt{m_t m_t} \\
\end{array}\right)\frac{\sqrt{2}}{vc_{\gamma^u} s_\beta}~,\\
\label{Yu-CS2}
{\bf Y}^{d1}&\simeq &\left(\begin{array}{ccc}
0 & 0 & 0 \\
0 & 0 & 0.3\sqrt{m_s m_b} \\
0&  2.9\sqrt{m_s m_b}  & 0 \\
\end{array}\right)\frac{\sqrt{2}}{vs_{\gamma^d} c_\beta} ~,~
{\bf Y}^{d2} \simeq\left(\begin{array}{ccc}
0 & 0 & 1.5\sqrt{m_d m_b}\\
0 & 0 & 0 \\
  -13\sqrt{m_d m_b} &0 & 0 \\
\end{array}\right)\frac{\sqrt{2}}{vs_{\gamma^d} c_\beta}~,\nn\\
{\bf Y}^{d3}&\simeq&\left(\begin{array}{ccc}
0 & 1.0\sqrt{m_d m_s} & 0\\
1.0\sqrt{m_d m_s} & 0 & 0 \\
0 &  0 & 0.9\sqrt{m_b m_b} \\
\end{array}\right)\frac{\sqrt{2}}{vc_{\gamma^d} c_\beta}~,
\label{Yd-CS2}
\ee}
where we have used the central values of the parameters given in (\ref{bestPII}). 
The large $(3,2)$ and $(3,1)$ elements of ${\bf Y}^{u1}$ and
${\bf Y}^{u2}$, respectively, can induce flavor changing
decay of the top quark, $t \to c+\bar{q}+q$ and $t \to u+\bar{q}+q$.
However, the rate is very  small.
The large $(1,2)$ and $(2,1)$ elements of ${\bf Y}^{u3}$ 
contribute dominantly to the mass deference of the neutral $D$ meson
$\Delta M_{D}\simeq 1.4\times 10^{-14}$ GeV.
From a similar estimate that we have done above  we find
\be
M_H & \gsim & 
\frac{\sqrt{2\cdot12\cdot 12 m_u m_c} }{vc_{\gamma^u} s_\beta}
\times 2.1\times 10^3~~\mbox{TeV}
\simeq \frac{8.3}{c_{\gamma^u} s_\beta}~\mbox{TeV}~.
\label{condMH2}
\ee
Here we have  demonstrated that the Cheng-Sher
mechanism is partially working in the $Q_6$ model.
In the lepton sector one can do the same calculations, but
it turned out that the constraints on the lower bound on $M_H$
are much weaker than (\ref{condMH2}) (see for instance \cite{Mondragon:2007af}).

\subsection{FCNCs and CP violations from 
the soft supersymmetry breaking (SSB)}
We work in the super CKM basis and 
use mass insertion parameters $(\delta_{ij}^\alpha)_{LL,RR,LR}
~(\alpha=u,d,e)$ 
to parameterize FCNCs and CP violations coming
from the SSB sector.
We start with the  mass eigenstates of the quarks
$u_{iL} \to u_{iL}' = (U_{uL})_{ij}u_{jL}~,~
u_{iR}^c \to u_{iR}^{c\prime} = (U_{uR})_{ij}^* u_{jR}^c$,
where $u_1'=u, u_2'=c, u_3'=t$,
and similarly for the other matter fermions.
Then we go to the super CKM basis for the matter fermions:
$\tilde{u}_{iL} \to \tilde{u}_{iL}^{s} = (U_{uL})_{ij}\tilde{u}_{jL}~,~
\tilde{u}_{iR} \to \tilde{u}_{iR}^{s} = (U_{uR})_{ij}\tilde{u}_{jR}~,~\mbox{etc}$.
In the super CKM basis, the squark mass matrices 
 become 
\be
{\bf \tilde{M}}_{u}^{s2}&=&
\!\left( \begin{array}{cc}m_{\tilde{u}_iL}m_{\tilde{u}_jL}[{\bf 1}_{ij}\!+\!(\delta_{ij}^u)_{LL}]
&(\Delta_{ii}^u)_{LR} {\bf 1}_{il} \!+\!
 m_{\tilde{u}_i L}m_{\tilde{u}_l R}(\delta_{il}^u)_{LR}
\cr
(\Delta_{kk}^u)^*_{LR}{\bf 1}_{kj} \!+\!
 m_{\tilde{u}_k L}m_{\tilde{u}_j R}(\delta_{jk}^{*u})_{LR}&
m_{\tilde{u}_k R}m_{\tilde{u}_l R}[ {\bf 1}_{kl}\!+\!(\delta_{kl}^u)_{RR} ]
\end{array} 
\right)~,
\label{Mup}
\ee
and 
\be
{\bf \tilde{M}}_{d}^{s2}&=&
\left( \begin{array}{cc}m_{\tilde{d}_iL}m_{\tilde{d}_jL}[{\bf 1}_{ij}+(\delta_{ij}^d)_{LL}]
&(\Delta_{ii}^d)_{LR} {\bf 1}_{il} \!+\!
 m_{\tilde{d}_i L}m_{\tilde{d}_l R}(\delta_{il}^d)_{LR}
\cr
(\Delta_{kk}^d)^*_{LR}{\bf 1}_{kj} \!+\!
 m_{\tilde{d}_k L}m_{\tilde{d}_j R}(\delta_{jk}^{*d})_{LR}& 
m_{\tilde{d}_k R}m_{\tilde{d}_l R}[ {\bf 1}_{kl}+(\delta_{kl}^d)_{RR} ]
\end{array} 
\right)~,\nn\\
\label{Mdp}
\ee
respectively, where \cite{Hall:1985dx}
\be
m_{u_i}{\bf 1}_{ij} &=& (U_{uL}^\dag {\bf m}_u U_{uR})_{ij}~,~
m_{d_i}{\bf 1}_{ij} = (U_{dL}^\dag {\bf m}_d U_{dR})_{ij}~,\\
m_{\tilde{u}_i L(R)}^2 
&=& (U_{uL(R)}^\dag {\bf \tilde{m}}^2_{Q (u)}U_{uL(R)})_{ii}+m_{u_i}^2
+c_{L(R)}^u M_Z^2~,\nn\\
m_{\tilde{d}_i L(R)}^2 &=& (U_{dL(R)}^\dag {\bf \tilde{m}}^2_{Q(d)} U_{dL(R)})_{ii}
+m_{d_i}^2
+c_{L(R)}^d M_Z^2~,\nn\\
(\delta_{ij}^{u(d)})_{LL} &=&[(U_{u(d)L}^\dag {\bf \tilde{m}}^2_Q U_{u(d)L})_{ij}
-(U_{u(d)L}^\dag {\bf \tilde{m}}^2_Q U_{u(d)L})_{ii}{\bf 1}_{ij}]
/m_{\tilde{u}_i (\tilde{d}_i)L}m_{\tilde{u}_j (\tilde{d}_j)L},\label{dLLRR}\\
(\delta_{ij}^{u(d)})_{RR} &=&[(U_{u(d)R}^\dag {\bf \tilde{m}}^2_{u(d)} U_{u(d)R})_{ij}
-(U_{u(d)R}^\dag {\bf \tilde{m}}^2_{u(d)}  U_{u(d)R})_{ii}{\bf 1}_{ij}]
/m_{\tilde{u}_i(\tilde{d}_i) R}m_{\tilde{u}_j (\tilde{d}_j)R},\nn\\
(\Delta_{ij}^u)_{LR} &=&
(U_{uL}^\dag)_{ik} \{
-\mu_{IJ}(Y_{kl}^{uI})^*<\!h^{d0}_J\!>+
A^{*u}_{kl}({\bf m}_u)_{kl}\} (U_{uR})_{lj}~,\nn\\
(\Delta_{ij}^d)_{LR} &=&
(U_{dL}^\dag)_{ik} \{ 
\mu_{JI}(Y_{kl}^{dI})<\!h^{u0}_J\!>+
A^{*d}_{kl}({\bf m}_d)_{kl} \}(U_{dR})_{lj}~,\label{DLR}\\
(\delta_{ij}^{u(d)})_{LR} &=&[(\Delta_{ij}^{u(d)})_{LR}-
(\Delta_{ii}^{u(d)})_{LR}{\bf 1}_{ij}]
/m_{\tilde{u}_i (\tilde{d}_i)L}m_{\tilde{u}_j (\tilde{d}_j)R}~.\label{dLR}
\ee
So far no assumption on  flavor symmetry is made, and
the squark mass matrices (\ref{Mup}) and (\ref{Mdp})  cover  the case of
more than one pairs of Higgs doublet supermultiplets.

If three generations of a family are put into
a one-dimensional and two-dimensional irreps of 
any  group, then 
the soft scalar mass matrix for the sfermions has always a diagonal form: 
\be
{\bf \tilde{m}^2}_{Q,L} =
{m}^2_{\tilde q,\tilde \ell} ~\mbox{diag.}~(a_L^{q,\ell} ,a_L^{q,\ell} ,b_L^{q,\ell} )~,~
{\bf \tilde{m}^2}_{\alpha} =
{m}^2_{\tilde q,\tilde \ell} 
~\mbox{diag.}~(a_R^{\alpha} ,a_R^{\alpha} ,b_R^{\alpha}  )
~~~(\alpha=u,d,e)~,
\label{scalarmass}
\ee
where ${m}_{\tilde q,\tilde \ell}$ denote the average of the  squark 
and slepton masses, respectively,  and $(a_{L(R)}, b_{L(R)})$ are
dimensionless free real parameters of $O(1)$.
Further, flavor symmetry imposes
the soft trilinear  coupling matrices ($A$ terms) to have the same structure as the
Yukawa coupling matrices.
They are real, because we impose CP invariance at
the Lagrangian level.
The imaginary parts of
$\delta^\alpha_{ij}$ and $\Delta^\alpha_{ij}$
contribute to CP violating processes.
Recall that the soft scalar mass matrices are real,
because they are diagonal.
The unitary matrices $U_{uL}$, etc.  are complex, and so 
$(\delta_{ij}^{\alpha})_{LL,RR}$ are
complex in general. Note that the unitary matrices have the form
$U=R P O$ in the case of ${\cal P}_I$
and 
$U= P R O$ in the case of ${\cal P}_{II}$
(see (\ref{Pud}) and (\ref{R})), respectively, 
where only the phase matrices $P$  are
complex. Since all the soft scalar mass matrices are diagonal
with the structure ${\bf \tilde{m}}^2 \sim \mbox{diag.} (a,a,b)$,
they commute with $P$ and $R$,
so that $(\delta_{ij}^{\alpha})_{LL,RR}$ have no
imaginary part.  
Further, the phase of the $A$ term contribution
to $(\Delta^{\alpha}_{ij})_{LR}$ in (\ref{DLR}) 
comes from the complex VEVs (see (\ref{VEV2})),
because CP is only spontaneously broken.
Therefore,
the $A$ term contribution has
the same phase structure as the corresponding fermion mass matrix
$ {\bf m}_\alpha$. That is, defining the $A$ parameters as
$h_{\gamma}^{\alpha} = A_{\gamma}^{\alpha} Y_{\gamma}^{\alpha}~~
(\alpha=u,d,e~,~\gamma=a,b,b',c)$,
where $h_{\gamma}^{\alpha}$ are the soft tri-linear couplings,
and the Yukawa couplings are defined in (\ref{Yuq})-(\ref{Yun}),
we find that, except for 
 $(\Delta^e_{ij})_{LR}$ in the  case of ${\cal P}_{II}$, 
 the $A$ term contributions
to $(\Delta^{\alpha}_{ij})_{LR}$ are real, too.
Consequently,
there is no CP violation originating from the SSB sector.
Only the $\mu$ term contributions
to $(\Delta^\alpha_{ij})_{LR}$ 
are complex.
The stringent constraints 
 coming from
the EDMs \cite{Gabbiani:1996hi,Abel:2001vy} 
are automatically satisfied in this way
of phase alignment,
except for the $\mu$ term contributions.
To suppress the contribution to the EDMs, therefore,
one should have relatively small $\mu$'s \cite{Babu:2009nn}.
This is general if there are more than one pair of Higgs doublet
supermultiplets.

The theoretical values of $(\delta_{ij}^\alpha)_{LL,RR,LR}$ for the present model 
are calculated below \footnote{We do not consider the RG running effects
of  the SSB parameters \cite{Barbieri:1995tw}.
See also \cite{Deppisch:2004fa}.},
where 
\be
\Delta a_{L}^{q,\ell} &=& a_{L}^{q,\ell}-b_{L}^{q,\ell},~
\Delta a_{R}^{\alpha} =a_{R}^{\alpha}-b_{R}^{\alpha},~
\tilde{A}_{\gamma}^{\alpha} = 
A_{\gamma}^{\alpha}/{m}_{\tilde q,\tilde \ell}~
~(\alpha=u,d,e~,~\gamma=a,b,b',c)~,
\label{deltaAt1}
 \ee
are introduced, and we assume that $ m_{\tilde{q}_i L,R}m_{\tilde{q}_i L,R}
=\delta_{ij} m_{\tilde{q}}^2~,~m_{\tilde{\ell}_i L,R}m_{\tilde{\ell}_i L,R}
=\delta_{ij} m_{\tilde{\ell}}^2$.
In Table 2, 
theoretical values of $\delta$'s
and  their experimental bounds 
\cite{Gabbiani:1996hi,Misiak:1997ei}
are summarized.
\be
\mbox{Im}(\delta^{d}_{11})_{LR}
&\simeq&10^{-4}\mbox{Im}\left[~
(-0.20 c_{\phi^d}+i 1.0s_{\phi^d})~\tilde{\mu}_+^d 
-(0.20 s_{\phi^d}+i 2.2c_{\phi^d})\tilde{\mu}_-^d+
0.05 ~ \tilde{\mu}_3^d~\right]\tilde{m}_{\tilde{q}}^{-1} ,\label{LRd11}\\
(\delta^{d}_{12})_{LR}
&\simeq&10^{-5}\left[~
~1.7\left( \tilde{A}_{a}^{d}-\tilde{A}_{b}^{d}-\tilde{A}_{b'}^{d}+ \tilde{A}_{c}^{d} \right)
\right.\nn\\
& -&\left.(8.6 c_{\phi^d}+i 1.1s_{\phi^d})~\tilde{\mu}_+^d 
-(8.6 s_{\phi^d}-i 1.1c_{\phi^d})\tilde{\mu}_-^d+
1.0 ~ \tilde{\mu}_3^d~\right]\tilde{m}_{\tilde{q}}^{-1} ,\label{LRd12}\\
(\delta^{d}_{21})_{LR}
&\simeq&10^{-4}\left[~
-0.16 \left( \tilde{A}_{a}^{d}-\tilde{A}_{b'}^{d} \right)
+0.19 \left( \tilde{A}_{b}^{d}-\tilde{A}_{c}^{d} \right)
\right.\nn\\
& +&\left.(0.78 c_{\phi^d}-i 3.8s_{\phi^d})~\tilde{\mu}_+^d 
+(0.78 s_{\phi^d}+i 3.8c_{\phi^d})\tilde{\mu}_-^d-
1.1 ~ \tilde{\mu}_3^d~\right]\tilde{m}_{\tilde{q}}^{-1} ,\nn\\
(\delta^{d}_{13})_{LR}
&\simeq&10^{-5}\left[~
-4.2 \left( \tilde{A}_{a}^{d}-\tilde{A}_{b}^{d} \right)
\right.\nn\\
& -&\left.(4.4 c_{\phi^d}+i 2.8s_{\phi^d})~\tilde{\mu}_+^d 
-(4.4 s_{\phi^d}-i 2.8c_{\phi^d})\tilde{\mu}_-^d+
0.42 ~ \tilde{\mu}_3^d~\right]\tilde{m}_{\tilde{q}}^{-1} ,\label{LRd13}\\
(\delta^{d}_{31})_{LR}
&\simeq&10^{-3}\left[~-0.40 \left( \tilde{A}_{a}^{d}-\tilde{A}_{b'}^{d} \right)
\right.\nn\\
& +&\left.(2.1 c_{\phi^d}-i 9.6s_{\phi^d})~\tilde{\mu}_+^d 
+(2.1 s_{\phi^d}+i 9.6c_{\phi^d})\tilde{\mu}_-^d-
0.004 ~ \tilde{\mu}_3^d~\right]\tilde{m}_{\tilde{q}}^{-1} ,\nn\\
(\delta^{d}_{23})_{LR}
&\simeq&10^{-4}\left[~
1.6 \tilde{A}_{a}^{d}-1.9 \tilde{A}_{b}^{d}
+0.3 \tilde{A}_{c}^{b'}
\right.\nn\\
& +&\left.(1.8 c_{\phi^d}-i 0.06s_{\phi^d})~\tilde{\mu}_+^d 
+(1.8 s_{\phi^d}+i 0.06c_{\phi^d})\tilde{\mu}_-^d+
0.03 ~ \tilde{\mu}_3^d~\right]\tilde{m}_{\tilde{q}}^{-1} ,\label{LRd23}\\
(\delta^{d}_{32})_{LR}
&\simeq&10^{-3}\left[~
-1.7 \left( \tilde{A}_{a}^{d}-\tilde{A}_{b}^{d} \right)
\right.\nn\\
& +&\left.(8.8 c_{\phi^d}+i 2.2s_{\phi^d})~\tilde{\mu}_+^d 
+(8.8 s_{\phi^d}-i 2.2c_{\phi^d})\tilde{\mu}_-^d+
0.02 ~ \tilde{\mu}_3^d~\right]\tilde{m}_{\tilde{q}}^{-1} ,\nn\\
\mbox{Im}(\delta^{u}_{11})_{LR}
&\simeq&3.5\times 10^{-4}\mbox{Im}(\tilde{\mu}_3^u)~
\tilde{m}_{\tilde{q}}^{-1} ~,\label{LRu11}\\
(\delta^{u}_{12})_{LR}
&\simeq&10^{-3}\left[~
-0.03\left(\tilde{A}_{a}^{u}-\tilde{A}_{b'}^{u}\right)
+0.03\left(\tilde{A}_{b}^{u}-\tilde{A}_{c}^{u}
\right)\right.\nn\\
& -&\left.(0.01 c_{\phi^u}-i 2.2s_{\phi^u})~\tilde{\mu}_+^u 
-(0.01 s_{\phi^u}+i 2.2c_{\phi^u})\tilde{\mu}_-^u+
0.6 ~ \tilde{\mu}_3^u~\right]\tilde{m}_{\tilde{q}}^{-1},\label{LRu12}\\
(\delta^{u}_{21})_{LR}
&\simeq&10^{-3}\left[~
-0.03\left(\tilde{A}_{a}^{u}-\tilde{A}_{b'}^{u}\right)
+0.03\left(\tilde{A}_{b}^{u}-\tilde{A}_{c}^{u}
\right)\right.\nn\\
&+&\left.(0.92 c_{\phi^u}+i 0.01s_{\phi^u})~\tilde{\mu}_+^u 
+(0.92 s_{\phi^u}-i 0.01c_{\phi^u})\tilde{\mu}_-^u-
1.6 ~ \tilde{\mu}_3^u~\right]\tilde{m}_{\tilde{q}}^{-1}
~,\nn
\ee
where
\be
\tilde{\mu}_+^u &=&
-\left( \frac{\mu_{11}}{m_{\tilde q}}\frac{c_{\phi^d}s_{\gamma^d}}{
s_{\gamma^u}}+\sqrt{2} ~ \frac{\mbox{Re} (\mu_{13})}{m_{\tilde q}}
\frac{c_{\gamma^d}}{s_{\gamma^u}}\right)\tan^{-1}\beta~,\nn\\
\tilde{\mu}_-^u &=&
\left( \frac{\mu_{11}}{m_{\tilde q}}\frac{s_{\phi^d}s_{\gamma^d}}{
s_{\gamma^u}}-\sqrt{2}  ~\frac{\mbox{Im} (\mu_{13})}{m_{\tilde q}}
\frac{c_{\gamma^d}}{s_{\gamma^u}}\right)~\tan^{-1}\beta~,\\
\tilde{\mu}_3^u &=&
-2~\frac{\mbox{Re}(\mu_{31} e^{-i \phi^d})}{m_{\tilde q}}
\frac{s_{\gamma^d}}{s_{\gamma^u}}~\tan^{-1}\beta~,
~\tilde{\mu}_3^d =
2~\frac{\mbox{Re}(\mu_{13} e^{-i \phi^u})}{m_{\tilde q}}
\frac{s_{\gamma^u}}{s_{\gamma^d}}~\tan\beta~,\nn\\
\tilde{\mu}_+^d &=&
\left( \frac{\mu_{11}}{m_{\tilde q}}\frac{c_{\phi^u}s_{\gamma^u}}{
s_{\gamma^d}}+\sqrt{2} ~ \frac{\mbox{Re} (\mu_{31})}{m_{\tilde q}}
\frac{c_{\gamma^u}}{s_{\gamma^d}}\right)~\tan\beta,
\tilde{\mu}_-^d =
\left( -\frac{\mu_{11}}{m_{\tilde q}}\frac{s_{\phi^u}s_{\gamma^u}}{
s_{\gamma^d}}+\sqrt{2}  ~\frac{\mbox{Im} (\mu_{31})}{m_{\tilde q}}
\frac{c_{\gamma^u}}{s_{\gamma^d}}\right)~\tan\beta~,\nn
\ee
and the angles $s_{\gamma^d}$, etc. are defined in  (\ref{cosgamma2})..
The $\mu$ term contributions above  are complex and so contribute to EDMs.
We see from from  (\ref{LRd11}) and  (\ref{LRu11})  that
the strong constraint 
$|\mbox{Im} (\delta^{u,d}_{11})'_{LR}| \lsim 10^{-6}$ from
the neutron EDM \cite{Gabbiani:1996hi,Abel:2001vy} can be satisfied if $\mu$'s are relatively small 
compared with the soft scalar masses \cite{Babu:2009nn,Kubo:2010mh}.
In contrast to this, $(\delta^{d}_{23})_{LR}$
and $(\delta^{d}_{32})_{LR}$, which contribute to $b\to s+\gamma$, are
well below the upper limit $10^{-2}\left( 500 ~\mbox{GeV}/m_{\tilde{q}} \right)^2$.
Further, the imaginary part of $(\delta^{d}_{23})_{LR}$
and $(\delta^{d}_{32})_{LR}$ contributes to the CP violation in the 
$B_s^0-\bar{B}_s^0$ mixing. Their size is small compared with
 $(\delta^{d}_{23})_{LL,RR}$ given in Table 2, which however 
 do not contribute to  CP violation because they are real at the tree-level.
 At the one-loop level they can be complex, and the size can be as large 
 as the tree-level size under certain circumstances \cite{Kubo:2010mh}.
With this (fine-tuned) one-loop contribution  the theoretical value of the  dimuon asymmetry
can  become close \cite{Kubo:2010mh}
to the order of the D0 measurement  \cite{Abazov:2010hv}.
This large CP violation has been toned down by CDF \cite{CDF:2011af} and than
by the resent measurement at  LHCb \cite{LHCb:2011aa} (see also \cite{Lenz:2012az}).
 Therefore, that fine tuning is no longer necessary.
Another interesting observation can be made on 
flavor violating decays of the top quark, i.e. $t \to u(c)+g$.
From $(\delta^u_{13,31,23,32})_{LR}$,
we find the ratio $B(t \to u+g)/B(t \to c+g)\sim 10^{-4}$,
which is consistent with the bound recently found by ATLAS \cite{Collaboration:2012gd}.

{\small  \begin{table}[thb]
\begin{center}
\begin{tabular}{|c||c|c|}
 \hline
 &  Exp. bound  & $Q_6$ Model $({\cal P}_{II})$\\
   \hline  \hline
$\sqrt{|\mbox{Re}(\delta^d_{12})^2_{LL,RR}|}$
& $4.0 \times 10^{-2} ~\tilde{m}_{\tilde{q}}$
& $(LL)4.0 \times 10^{-4} \Delta a_L^q,(RR)3.3 \times 10^{-2}\Delta a_R^d$
\\ \hline
 $\sqrt{|\mbox{Re}(\delta^d_{12})_{LL}(\delta^d_{12})_{RR}|}$
& $2.8 \times 10^{-3} ~\tilde{m}_{\tilde{q}}$
&$3.6 \times 10^{-3}\sqrt{\Delta a_L^q \Delta a_R^d}$
\\ \hline
$\sqrt{|\mbox{Re}(\delta^d_{12})^2_{LR}|}$
 & $4.4 \times 10^{-3} ~\tilde{m}_{\tilde{q}}$
& Eq.~(\ref{LRd12})
\\ \hline
 $\sqrt{|\mbox{Re}(\delta^d_{13})^2_{LL,RR}|}$
 & $9.8 \times 10^{-2} ~\tilde{m}_{\tilde{q}}$
&$(LL)1.0\times 10^{-2}\Delta a_L^q,(RR)8.0 \times 10^{-2}\Delta a_R^d$
\\ \hline
$\sqrt{|\mbox{Re}(\delta^d_{13})_{LL}(\delta^d_{13})_{RR}|}$
&  $1.8 \times 10^{-2} ~\tilde{m}_{\tilde{q}}$
&$2.8 \times 10^{-2}\sqrt{\Delta a_L^q \Delta a_R^d}$
\\ \hline
 $\sqrt{|\mbox{Re}(\delta^d_{13})^2_{LR}|}$
& $3.3 \times 10^{-2} ~\tilde{m}_{\tilde{q}}$
& Eq.~(\ref{LRd13})
\\ \hline
$\sqrt{|\mbox{Re}(\delta^u_{12})^2_{LL,RR}|}$
& $1.5 \times 10^{-1} ~\tilde{m}_{\tilde{q}}$
&$(LL)2.9 \times 10^{-8}\Delta a_L^q,(RR)4.3 \times 10^{-1}\Delta a_R^u$
\\ \hline
 $\sqrt{|\mbox{Re}(\delta^u_{12})_{LL}(\delta^u_{12})_{RR}|}$
& $1.7 \times 10^{-2} ~\tilde{m}_{\tilde{q}}$
&$1.1 \times 10^{-4}\sqrt{\Delta a_L^q \Delta a_R^u}$
\\ \hline
$\sqrt{|\mbox{Re}(\delta^u_{12})^2_{LR}|}$
& $3.1 \times 10^{-2} ~\tilde{m}_{\tilde{q}}$
& Eq.~(\ref{LRu12})
\\ \hline
$|(\delta^d_{23})_{LL,RR}|$
 & $8.2~ \tilde{m}_{\tilde{q}}^2$
 &$(LL)3.9 \times 10^{-2}\Delta a_L^q,(RR)3.5 \times 10^{-1}\Delta a_R^d$
 \\ \hline
 $|(\delta^d_{23})_{LR}|$
& $1.6 \times 10^{-2} ~\tilde{m}_{\tilde{q}}^2$
 &Eq.~(\ref{LRd23})\\ 
\hline
 $|\mbox{Im}(\delta^d_{11})_{LR}|$
& $3.0 \times 10^{-6} ~\tilde{m}_{\tilde{q}}$
 &Eq.~(\ref{LRd11})
\\ 
\hline
 $|\mbox{Im}(\delta^u_{11})_{LR}|$
& $5.9 \times 10^{-6} ~\tilde{m}_{\tilde{q}}$
 &Eq.~(\ref{LRu11})
\\ 
\hline
\end{tabular}
\caption{\footnotesize{Experimental bounds on  $(\delta_{ij}^{u,d})_{LL,RR,LR}$
 \cite{Gabbiani:1996hi,Misiak:1997ei} and their theoretical values  in the
quark sector of the $Q_6$ model for ${\cal P}_{II}$
(see also Kobayashi {\it et al.} in \cite{Kobayashi:2003fh}),
 where 
$\tilde{m}_{\tilde{q}}$ denotes
$m_{\tilde{q}} /500$ GeV, and $\Delta a_{L,R}$
 and $\tilde A$ are given 
in (\ref{deltaAt1}).}}
\end{center}
\end{table}}
Now we come to the lepton sector.
\begin{figure}
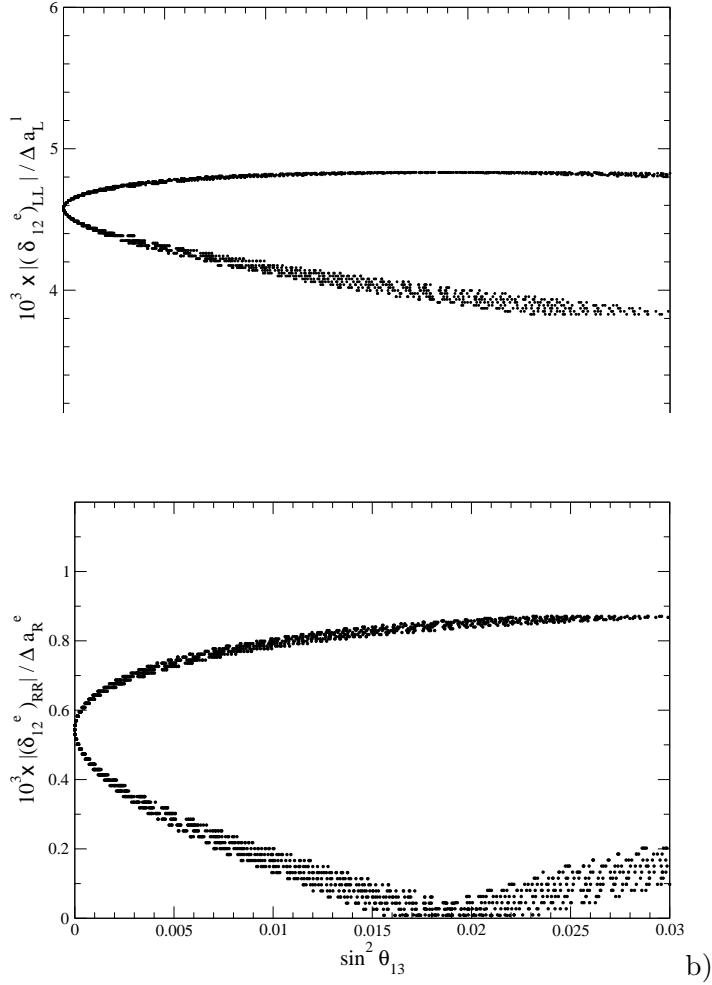

\includegraphics[width=9cm]{v13-dLL12.eps}a)
\hfil
\includegraphics[width=9cm]{v13-dRR12.eps}b)
\caption{ \footnotesize{$|(\delta^e_{12})_{LL}|/\Delta a_L^\ell$ against $\sin^2\theta_{13}$ (a)
and
$|(\delta^e_{12})_{RR}|/\Delta a_R^e$ (b).
The  parameter space is the same as for Fig.~\ref{prediction-n-1}.}}
\label{fig4}
\end{figure}
\begin{figure}
\includegraphics*[width=12cm]{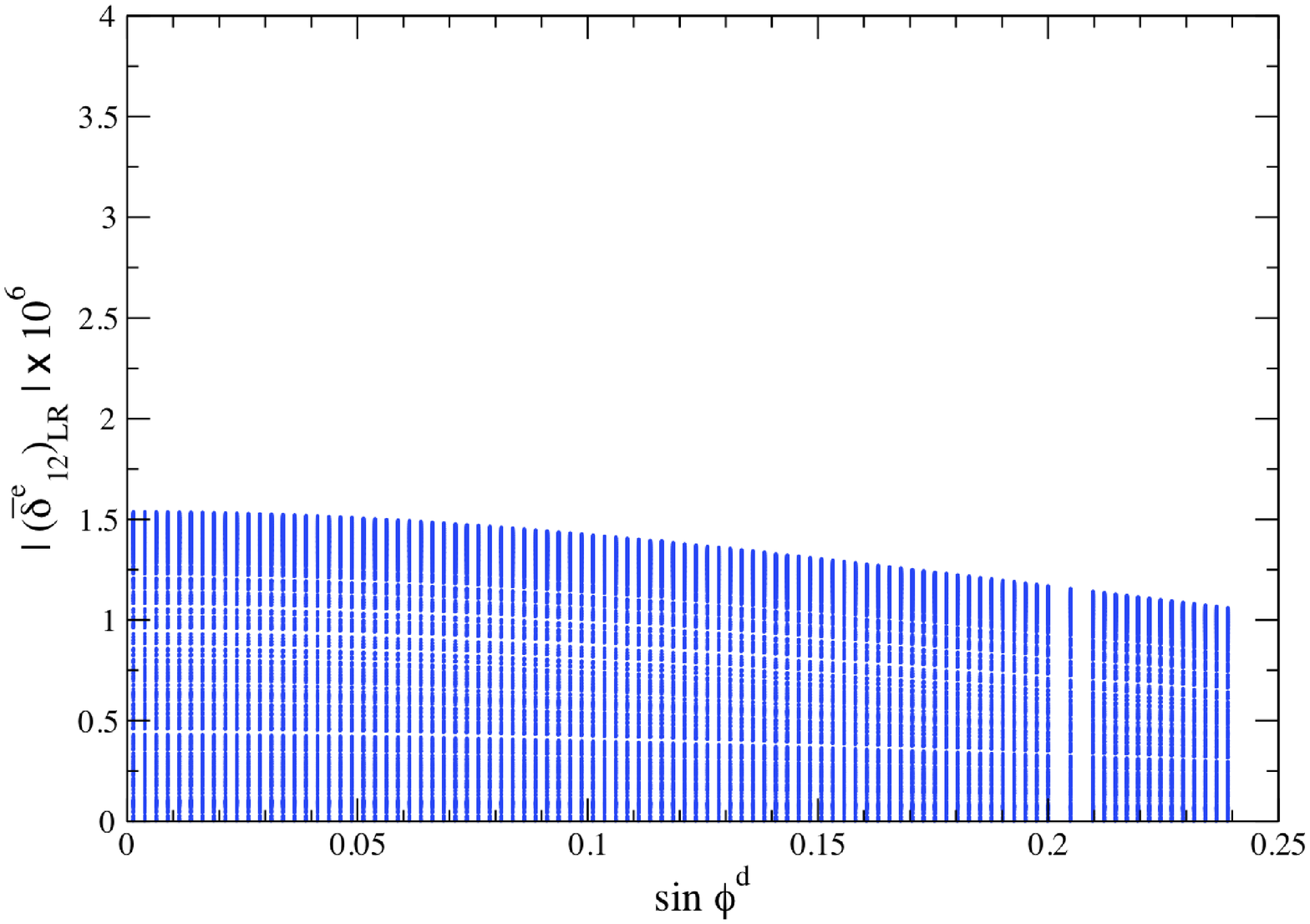}
\caption{\label{fig5}\footnotesize
$ (\bar{\delta}^e_{12})_{LR} $ against $ \sin\phi^d $, 
where the dimensionless parameters $\tilde{A}_\gamma^e$ are varied from $-5$ to $5$. $(\bar{\delta}^e_{ij})_{LR}$ is defined in (\ref{deltabar}).}
\end{figure}
As it is noticed, the phase alignment does not work
for the lepton sector in the case of ${\cal P}_{II}$.
There are two phases, $\phi^u$ and $\phi^d$,
 that enter into the CKM and MNS matrices.
One combination $\phi=\phi^u-\phi^d$ is basically fixed 
around $0.0838$ (see (\ref{bestPII}))  to produce the correct
Kobayashi-Maskawa CP phase.
The other would be fixed if CP violation in the neutrino sector
is precisely measured.
Then the phase that enters into the soft mass insertions
would be completely fixed.
In other words we can express the mass insertions as a function
of $\sin^2\theta_{13}$, because in the case of ${\cal P}_{II}$
the CP phase $\delta_{CP}$ and $\sin^2\theta_{13}$
are intimately related.
This is done in Fig.~\ref{fig4}.
We plot 
$|(\delta^e_{12})_{LL}|/\Delta a_L^\ell$ 
and $|(\delta^e_{12})_{RR}|/\Delta a_L^\ell$ against
$\sin^2\theta_{13}$, respectively.
We see from  Fig.~\ref{fig4}  that
$|(\delta^e_{12})_{LL}|/\Delta a_L^\ell$ does not change very much
as $\sin^2\theta_{13}$ varies, while 
$|(\delta^e_{12})_{RR}|/\Delta a_R^\ell$ depends 
significantly on $\sin^2\theta_{13}$.
Unfortunately, there is no common range of
 $\sin^2\theta_{13}$ such that the both insertions
 become small.
 The experimental bounds are \cite{Gabbiani:1996hi,Misiak:1997ei} 
 \footnote{
The recent experimental bounds  $B(\mu\to e + \gamma) \lsim 2.4\times 10^{-12}$
\cite{Adam:2011ch} and $B(\tau \to \mu (e)+\gamma)
\lsim 4.4 (3.3) \times 10^{-8}$ \cite{Hayasaka:2010et} are included.}: 
 \begin{equation}
 |(\delta_{12}^e)_{LL}| \lsim  1.8\times 10^{-4} 
\tilde{m}_{\tilde{\ell}}^2~,~
  |(\delta_{12}^e)_{RR}| \lsim  3.6\times 10^{-4} 
\tilde{m}_{\tilde{\ell}}^2~,~
  |(\delta_{12}^e)_{LR}| \lsim  3.3\times 10^{-6} 
 \tilde{m}_{\tilde{\ell}}^2~,\label{LRe12}
 \end{equation}
 where $\tilde{m}_{\tilde{\ell}}=m_{\tilde{\ell}}/300 \mbox{GeV}$. 
  In Fig.~\ref{fig5} we plot
$ |(\bar{\delta}^{e}_{12})_{LR} |$
against $\sin\phi^d$ with $\phi$ fixed at 
$0.08375$, where we varied the dimensionless parameters
$\tilde{A}^{e}_{\gamma}$ (defined in (\ref{deltaAt1}) ) from $-5$ to $5$, and
 $ |(\bar{\delta}^{e}_{12})_{LR} |$ is defined as
   \be
  (\bar{\delta}^{e}_{12})_{LR} &=&
    (\delta^{e}_{12})_{LR}~\tilde{m}_{\tilde{\ell}}^{-1}~
       /( \sum_{\gamma=a,b,b',c}|\tilde{A}^{e}_{\gamma}| )~.
    \label{deltabar}
  \ee
We see from Fig.~\ref{fig5} that the size of
 $ (\delta^{e}_{12})_{LR} $ is of order of the
 experimental bound (\ref{LRe12}), and that it has no strong
 dependence of $\sin\phi^d$ and hence of  $\sin^2\theta_{13}$.

We see thus from Tables 2 and Figs.~\ref{fig4}and ~\ref{fig5}  that
 $Q_6$ flavor symmetry can well soften the FCNC problem of the 
SSB sector.
The recent LHC data suggest that the superpartner masses are 
heavier than those assumed here \cite{Chatrchyan:2011zy,Aad:2011ib}. If this is the case
the constraints are better satisfied.

\section{Summary}

Obviously, the flavor problem is one of the most
difficult problems. 
So far we do not have a magic bullet to solve this problem.
Non-abelian flavor symmetry can soften the problem and
therefore might give a hint to a solution. 
In this contribution I restricted myself 
to  renormalizable models of flavor with
a non-abelian discrete flavor symmetry,
which is at most softly broken at some tera scale.
This setting without supersymmetry may lead
to problems because of FCNCs and of a fine tuning in the scalar
potential. The fine tuning problem may be softened
in supersymmetric models.
In  softly broken supersymmetric theories,
the soft breaking terms of flavor symmetry
can be classified, which I discussed in a very general form.

To point out basic features of supersymmetric models of flavor
with multiple Higgs doublets I considered a specific model based
on the finite flavor group $Q_6$ \cite{Babu:2004tn}.
To make spontaneous breaking of CP possible,
the Higgs sector was further extended \cite{Babu:2011mv}
so as to include  a certain set of SM singlet Higgs multiplets.
With them the upper bound of the lightest Higgs mass upper bound can  slightly
increase.
In this   SM singlet Higgs sector, 
flavor symmetry and  CP 
are spontaneously broken at a tera scale,
which is very close to supersymmetry breaking scale.
This may give a hint that 
flavor symmetry, CP and supersymmetry 
are broken within the same sector.

The $Q_6$ model of flavor yields
interesting predictions for the CKM and MNS mixing parameters, 
which compare very well with experimental data.
Large B-factories \cite{Adeva:2009ny,Aushev:2010bq}
with more precise determination of
quark masses \cite{Colangelo:2010et}
will be able to test the predictions in the quark sector \cite{Araki:2008rn}.
In the lepton sector,  the predictions for the ground state ${\cal P}_{II}$
are particularly  interesting, because  the neutrino mass hierarchy is inverted
with nearly maximal CP violation along with
nearly maximal mixing of atmospheric neutrinos:
The CP phase, the average neutrino mass $<\!m_{ee}\!>$,
and $\sin^2\theta_{23}$ can be expressed as a function of 
$\sin^2\theta_{13}$. These predictions can also be tested
by the  future neutrino experiments
(see \cite{Ardellier:2006mn} and (\cite{Elliott:2012sp}).

Non-abelian flavor symmetry can not only relate
the fermion masses and mixing parameters, but also
 suppress FCNCs.
Flavor symmetry embeds  the  Cheng-Sher mechanism 
automatically,
which suppresses the tree-level FCNCs, making the flavor symmetry scale decrease down to
 tera scale, and  
 suppress FCNCs coming from the soft breaking of supersymmetry as well.
The CP problem of softly broken supersymmetric
theories is solved  by virtue of spontaneous CP violation.
This implies that the soft supersymmetry breaking  parameters
such as the gluino mass are all real.  The phase of the trilinear
soft breaking $A$ terms,
since they arise spontaneously, will align with the phases in the fermion
mass matrices. A self-alignment mechanism is going on.
Thus the $A$ terms do not generate CP violation. 
Thanks to flavor symmetry, the mass insertion parameters
$(\delta_{ij}^\alpha)_{LL,RR,LR}$ depend on only few independent
parameters. Their ratio, e.g., $(\delta_{12}^e)_{LL}/(\delta_{13}^e)_{RR}$,
is  fixed, which is a unique prediction.
Therefore,  flavor violations
are controlled by flavor symmetry.
Thus new observations of   flavor and CP violations
can also test flavor symmetry at tera scale \cite{Calibbi:2011dn}.

\section*{Acknowledgments}  
I thank Jose  Valle, Stefano Morisi
and Claudia Hagedorn
for inviting me to a contribution
to the special issue
 on "Flavor Symmetries and Neutrino Oscillations".
 I also thank T.~Araki, K.S.~Babu, E.~Itou,  
Y.~Kaburaki, Y.~Kajiyama, K.~Kawashima, N.~Kifune,  
 T.~Kobayashi, K.~Konya, A.~Lenz, 
A.~Mondragon, M.~Mondragon, E.~Rodriguez-Jauregui, 
 H.~Okada, F.~Sakamaki, 
H.~Terao
for exiting collaborations. Many results presented here have been obtained
with them.
The work of JK is partially supported by a Grant-in-Aid for
Scientific Research (C) from Japan Society for the Promotion of Science (No. 22540271).

\end{document}